\RequirePackage{lineno}
\documentclass[aps,prd,twocolumn,superscriptaddress,preprintnumbers,amsmath,amssymb,floatfix]{revtex4-1}

\usepackage{graphicx} 
\usepackage{dcolumn}  
\usepackage{graphicx} 
\usepackage{dcolumn}  
\usepackage{xspace}
\usepackage{cleveref}
\usepackage{amsmath} 
\usepackage{subfigure}
\usepackage{booktabs}
\usepackage{longtable}
\usepackage{dcolumn}
\graphicspath{{ps}}

\newcommand*\patchAmsMathEnvironmentForLineno[1]{%
  \expandafter\let\csname old#1\expandafter\endcsname\csname #1\endcsname
  \expandafter\let\csname oldend#1\expandafter\endcsname\csname end#1\endcsname
  \renewenvironment{#1}%
     {\linenomath\csname old#1\endcsname}%
     {\csname oldend#1\endcsname\endlinenomath}}%
\newcommand*\patchBothAmsMathEnvironmentsForLineno[1]{%
  \patchAmsMathEnvironmentForLineno{#1}%
  \patchAmsMathEnvironmentForLineno{#1*}}%
\AtBeginDocument{%
\patchBothAmsMathEnvironmentsForLineno{equation}%
\patchBothAmsMathEnvironmentsForLineno{align}%
\patchBothAmsMathEnvironmentsForLineno{flalign}%
\patchBothAmsMathEnvironmentsForLineno{alignat}%
\patchBothAmsMathEnvironmentsForLineno{gather}%
\patchBothAmsMathEnvironmentsForLineno{multline}%
}

 \makeatletter
 \DeclareRobustCommand*{\bfseries}{%
   \not@math@alphabet\bfseries\mathbf
   \fontseries\bfdefault\selectfont
   \boldmath
 }

\begin{document}

\preprint{\vbox{ \hbox{   }
						\hbox{BELLE-CONF-1603}
%
}}

\title{ \quad\\[0.5cm] Angular Analysis of \bkllzero}
\newcommand{\phit}{\textless phi-t\textgreater\textsuperscript\textregistered\,}
\newcommand{\neurobayes}{NeuroBayes\textsuperscript\textregistered\,}

%
\newcommand{\yvs}{$\Upsilon(5S)$ }
\newcommand{\sm}{Standard Model\xspace}
\newcommand{\SM}{Standard Model\xspace}
\newcommand{\NP}{new physics\xspace}
\newcommand{\B}{$B$ meson\xspace}
\newcommand{\Bs}{$B$ mesons\xspace}
\newcommand{\Bjpsi}{$B\to K^\ast J/\psi$\xspace}
\newcommand{\Bpsi}{$B\to K^\ast \psi(2S)$\xspace}
\newcommand{\Bto}[1]{
	\ifnum #1 = 521343 \ensuremath{B^+ \to K^+ e^+ e^- \xspace}\fi
	\ifnum #1 = 521347 \ensuremath{B^+ \to K^+ \mu^+ \mu^- \xspace}\fi
	\ifnum #1 = 521345 \ensuremath{B^+ \to K^\ast(892)^+ e^+ e^- \xspace}\fi
	\ifnum #1 = 521349 \ensuremath{B^+ \to K^\ast(892)^+ \mu^+ \mu^- \xspace}\fi
	\ifnum #1 = 511332 \ensuremath{B^0 \to K^0 e^+ e^- \xspace}\fi
	\ifnum #1 = 511336 \ensuremath{B^0 \to K^0 \mu^+ \mu^- \xspace}\fi
	\ifnum #1 = 511335 \ensuremath{B^0 \to K^\ast(892)^0 e^+ e^- \xspace}\fi
	\ifnum #1 = 511339 \ensuremath{B^0 \to K^\ast(892)^0  \mu^+ \mu^- \xspace}\fi
}
\newcommand{\Kto}[1]{
	\ifnum #1 = 313421 \ensuremath{K^{\ast 0} \to K_S \pi^0}\fi
	\ifnum #1 = 313532 \ensuremath{K^{\ast 0} \to K^+ \pi^-}\fi
	\ifnum #1 = 323432 \ensuremath{K^{\ast +} \to K^+ \pi^0}\fi
	\ifnum #1 = 323521 \ensuremath{K^{\ast +} \to K_S \pi^+}\fi
}
\newcommand{\mbc}{\ensuremath{M_\mathrm{bc}}\xspace}
\newcommand{\de}{\ensuremath{\Delta E}\xspace}
\newcommand{\ctl}{$\cos\theta_\ell$\xspace}
\newcommand{\ctk}{$\cos\theta_K$\xspace}
\newcommand{\ph}{$\phi$\xspace}
\newcommand{\nbout}{$NB_{out}$}
\newcommand*{\factor}{0.95}
\newcommand*{\factorh}{0.48}
\newcommand*{\factort}{0.315}
\newcommand*{\factorq}{0.24}
\newcommand{\yfs}{$\Upsilon(4S)$\xspace}
\newcommand{\bkll}{$B \to K^{(\ast)} \ell^+ \ell^-$\xspace}
\newcommand{\bkllzero}{\ensuremath{B^0 \to K^\ast(892)^0  \ell^+ \ell^-\xspace}}
\newcommand{\bsll}{$b \to s \ell^+ \ell^-$\xspace}
\newcommand{\bktt}{\ensuremath{B^+ \to K^+ \tau^+ \tau^-}\xspace}
\newcommand{\bkee}{\ensuremath{B^+ \to K^+ e^+ e^-}\xspace}
\newcommand{\bkttp}{\ensuremath{B^+ \to K^+ \tau^+ \tau^-}\xspace}
\newcommand{\bkllp}{$B \to K^{(\ast)} \ell^+ \ell^-$\xspace}
\newcommand{\bmn}{$B$ meson\xspace}
\newcommand{\bms}{$B$ mesons\xspace}
\newcommand{\kast}{$K^\ast$\xspace}
\newcommand{\bc}{$B$ candidate\xspace}
\newcommand{\cms}{center-of-mass\xspace}
\newcommand{\nbb}{772 million\xspace}
\newcommand{\eecl}{\ensuremath{E_{ECL}}\xspace}
\newcommand{\btag}{$B_{tag}$\xspace}
\newcommand{\upperlimm}{${\cal B}(B^+ \to K^+ \tau^+ \tau^-)< 3.17\times 10^{-4}$\xspace}
\newcommand{\cutbdt}{$c_{BDT}=0.553$\xspace}
\newcommand{\cuteecl}{$c_{E}=0.206~\mathrm{GeV}$\xspace}
\newcommand{\nbkg}{$8.83$\xspace}
\newcommand{\effsig}{$\epsilon_s=2.709\times10^{-5}$\xspace}
\newcommand{\upl}[1]{\ensuremath{{\cal B}(B^+ \to K^+ \tau^+ \tau^-)< #1}}
\newcommand{\upll}[2]{\ensuremath{{\cal B}(B^+ \to K^+ \tau^+ \tau^-)< #1 \times 10^{- #2}}}
\newcommand{\uplll}[3]{\ensuremath{{\cal B}(#1)< #2 \times 10^{- #3}}}
\newcommand{\upllll}[4]{\ensuremath{{\cal B}(#1)^{#2}< #3 \times 10^{- #4}}}
\newcommand{\br}[4]{\ensuremath{{\cal B}(#1)^{#2}= #3 \times 10^{- #4}}}
\newcommand*\diff{\mathop{}\!\mathrm{d}}
\newcommand*\Diff[1]{\mathop{}\!\mathrm{d^#1}}
\newcommand{\gevsq}{\ensuremath{~\mathrm{GeV}^2/c^4}\xspace}
\newcommand{\fitcomponents}{Combinatorial (dashed blue), signal (red filled) and total (solid) fit distributions are superimposed on the data points.}
\newcommand{\fitprojectioncaption}[2]{Projections for the fit result of $P_{#1}'$ in bin #2.  Fit to the \mbc sideband for the determination of the background shape (top) and signal region (bottom) are displayed. \fitcomponents}
\newcommand{\ckmt}[2]{\ensuremath{V_{#1}^{\;}V_{#2}^\ast}\xspace}
\noaffiliation
\affiliation{Aligarh Muslim University, Aligarh 202002}
\affiliation{University of the Basque Country UPV/EHU, 48080 Bilbao}
\affiliation{Beihang University, Beijing 100191}
\affiliation{University of Bonn, 53115 Bonn}
\affiliation{Budker Institute of Nuclear Physics SB RAS, Novosibirsk 630090}
\affiliation{Faculty of Mathematics and Physics, Charles University, 121 16 Prague}
\affiliation{Chiba University, Chiba 263-8522}
\affiliation{Chonnam National University, Kwangju 660-701}
\affiliation{University of Cincinnati, Cincinnati, Ohio 45221}
\affiliation{Deutsches Elektronen--Synchrotron, 22607 Hamburg}
\affiliation{University of Florida, Gainesville, Florida 32611}
\affiliation{Department of Physics, Fu Jen Catholic University, Taipei 24205}
\affiliation{Justus-Liebig-Universit\"at Gie\ss{}en, 35392 Gie\ss{}en}
\affiliation{Gifu University, Gifu 501-1193}
\affiliation{II. Physikalisches Institut, Georg-August-Universit\"at G\"ottingen, 37073 G\"ottingen}
\affiliation{SOKENDAI (The Graduate University for Advanced Studies), Hayama 240-0193}
\affiliation{Gyeongsang National University, Chinju 660-701}
\affiliation{Hanyang University, Seoul 133-791}
\affiliation{University of Hawaii, Honolulu, Hawaii 96822}
\affiliation{High Energy Accelerator Research Organization (KEK), Tsukuba 305-0801}
\affiliation{Hiroshima Institute of Technology, Hiroshima 731-5193}
\affiliation{IKERBASQUE, Basque Foundation for Science, 48013 Bilbao}
\affiliation{University of Illinois at Urbana-Champaign, Urbana, Illinois 61801}
\affiliation{Indian Institute of Science Education and Research Mohali, SAS Nagar, 140306}
\affiliation{Indian Institute of Technology Bhubaneswar, Satya Nagar 751007}
\affiliation{Indian Institute of Technology Guwahati, Assam 781039}
\affiliation{Indian Institute of Technology Madras, Chennai 600036}
\affiliation{Indiana University, Bloomington, Indiana 47408}
\affiliation{Institute of High Energy Physics, Chinese Academy of Sciences, Beijing 100049}
\affiliation{Institute of High Energy Physics, Vienna 1050}
\affiliation{Institute for High Energy Physics, Protvino 142281}
\affiliation{Institute of Mathematical Sciences, Chennai 600113}
\affiliation{INFN - Sezione di Torino, 10125 Torino}
\affiliation{J. Stefan Institute, 1000 Ljubljana}
\affiliation{Kanagawa University, Yokohama 221-8686}
\affiliation{Institut f\"ur Experimentelle Kernphysik, Karlsruher Institut f\"ur Technologie, 76131 Karlsruhe}
\affiliation{Kavli Institute for the Physics and Mathematics of the Universe (WPI), University of Tokyo, Kashiwa 277-8583}
\affiliation{Kennesaw State University, Kennesaw, Georgia 30144}
\affiliation{King Abdulaziz City for Science and Technology, Riyadh 11442}
\affiliation{Department of Physics, Faculty of Science, King Abdulaziz University, Jeddah 21589}
\affiliation{Korea Institute of Science and Technology Information, Daejeon 305-806}
\affiliation{Korea University, Seoul 136-713}
\affiliation{Kyoto University, Kyoto 606-8502}
\affiliation{Kyungpook National University, Daegu 702-701}
\affiliation{\'Ecole Polytechnique F\'ed\'erale de Lausanne (EPFL), Lausanne 1015}
\affiliation{P.N. Lebedev Physical Institute of the Russian Academy of Sciences, Moscow 119991}
\affiliation{Faculty of Mathematics and Physics, University of Ljubljana, 1000 Ljubljana}
\affiliation{Ludwig Maximilians University, 80539 Munich}
\affiliation{Luther College, Decorah, Iowa 52101}
\affiliation{University of Maribor, 2000 Maribor}
\affiliation{Max-Planck-Institut f\"ur Physik, 80805 M\"unchen}
\affiliation{School of Physics, University of Melbourne, Victoria 3010}
\affiliation{Middle East Technical University, 06531 Ankara}
\affiliation{University of Miyazaki, Miyazaki 889-2192}
\affiliation{Moscow Physical Engineering Institute, Moscow 115409}
\affiliation{Moscow Institute of Physics and Technology, Moscow Region 141700}
\affiliation{Graduate School of Science, Nagoya University, Nagoya 464-8602}
\affiliation{Kobayashi-Maskawa Institute, Nagoya University, Nagoya 464-8602}
\affiliation{Nara University of Education, Nara 630-8528}
\affiliation{Nara Women's University, Nara 630-8506}
\affiliation{National Central University, Chung-li 32054}
\affiliation{National United University, Miao Li 36003}
\affiliation{Department of Physics, National Taiwan University, Taipei 10617}
\affiliation{H. Niewodniczanski Institute of Nuclear Physics, Krakow 31-342}
\affiliation{Nippon Dental University, Niigata 951-8580}
\affiliation{Niigata University, Niigata 950-2181}
\affiliation{University of Nova Gorica, 5000 Nova Gorica}
\affiliation{Novosibirsk State University, Novosibirsk 630090}
\affiliation{Osaka City University, Osaka 558-8585}
\affiliation{Osaka University, Osaka 565-0871}
\affiliation{Pacific Northwest National Laboratory, Richland, Washington 99352}
\affiliation{Panjab University, Chandigarh 160014}
\affiliation{Peking University, Beijing 100871}
\affiliation{University of Pittsburgh, Pittsburgh, Pennsylvania 15260}
\affiliation{Punjab Agricultural University, Ludhiana 141004}
\affiliation{Research Center for Electron Photon Science, Tohoku University, Sendai 980-8578}
\affiliation{Research Center for Nuclear Physics, Osaka University, Osaka 567-0047}
\affiliation{RIKEN BNL Research Center, Upton, New York 11973}
\affiliation{Saga University, Saga 840-8502}
\affiliation{University of Science and Technology of China, Hefei 230026}
\affiliation{Seoul National University, Seoul 151-742}
\affiliation{Shinshu University, Nagano 390-8621}
\affiliation{Showa Pharmaceutical University, Tokyo 194-8543}
\affiliation{Soongsil University, Seoul 156-743}
\affiliation{University of South Carolina, Columbia, South Carolina 29208}
\affiliation{Sungkyunkwan University, Suwon 440-746}
\affiliation{School of Physics, University of Sydney, New South Wales 2006}
\affiliation{Department of Physics, Faculty of Science, University of Tabuk, Tabuk 71451}
\affiliation{Tata Institute of Fundamental Research, Mumbai 400005}
\affiliation{Excellence Cluster Universe, Technische Universit\"at M\"unchen, 85748 Garching}
\affiliation{Department of Physics, Technische Universit\"at M\"unchen, 85748 Garching}
\affiliation{Toho University, Funabashi 274-8510}
\affiliation{Tohoku Gakuin University, Tagajo 985-8537}
\affiliation{Department of Physics, Tohoku University, Sendai 980-8578}
\affiliation{Earthquake Research Institute, University of Tokyo, Tokyo 113-0032}
\affiliation{Department of Physics, University of Tokyo, Tokyo 113-0033}
\affiliation{Tokyo Institute of Technology, Tokyo 152-8550}
\affiliation{Tokyo Metropolitan University, Tokyo 192-0397}
\affiliation{Tokyo University of Agriculture and Technology, Tokyo 184-8588}
\affiliation{University of Torino, 10124 Torino}
\affiliation{Toyama National College of Maritime Technology, Toyama 933-0293}
\affiliation{Utkal University, Bhubaneswar 751004}
\affiliation{Virginia Polytechnic Institute and State University, Blacksburg, Virginia 24061}
\affiliation{Wayne State University, Detroit, Michigan 48202}
\affiliation{Yamagata University, Yamagata 990-8560}
\affiliation{Yonsei University, Seoul 120-749}
  \author{A.~Abdesselam}\affiliation{Department of Physics, Faculty of Science, University of Tabuk, Tabuk 71451} 
  \author{I.~Adachi}\affiliation{High Energy Accelerator Research Organization (KEK), Tsukuba 305-0801}\affiliation{SOKENDAI (The Graduate University for Advanced Studies), Hayama 240-0193} 
  \author{K.~Adamczyk}\affiliation{H. Niewodniczanski Institute of Nuclear Physics, Krakow 31-342} 
  \author{H.~Aihara}\affiliation{Department of Physics, University of Tokyo, Tokyo 113-0033} 
  \author{S.~Al~Said}\affiliation{Department of Physics, Faculty of Science, University of Tabuk, Tabuk 71451}\affiliation{Department of Physics, Faculty of Science, King Abdulaziz University, Jeddah 21589} 
  \author{K.~Arinstein}\affiliation{Budker Institute of Nuclear Physics SB RAS, Novosibirsk 630090}\affiliation{Novosibirsk State University, Novosibirsk 630090} 
  \author{Y.~Arita}\affiliation{Graduate School of Science, Nagoya University, Nagoya 464-8602} 
  \author{D.~M.~Asner}\affiliation{Pacific Northwest National Laboratory, Richland, Washington 99352} 
  \author{T.~Aso}\affiliation{Toyama National College of Maritime Technology, Toyama 933-0293} 
  \author{H.~Atmacan}\affiliation{Middle East Technical University, 06531 Ankara} 
% ...
\collaboration{The Belle Collaboration}

\noaffiliation

\begin{abstract}
We present a measurement of angular observables, $P_4'$, $P_5'$, $P_6'$, $P_8'$,  in the decay  $B^0 \to K^\ast(892)^0  \ell^+ \ell^-$, where $\ell^+\ell^-$ is either $e^+e^-$ or $\mu^+\mu^-$.
The analysis is performed on a data sample corresponding to an integrated luminosity of $711~\mathrm{fb}^{-1}$  containing $772\times 10^{6}$ $B\bar B$ pairs, collected at the \yfs resonance with the Belle detector at the asymmetric-energy $e^+e^-$ collider KEKB.
Four angular observables, $P_{4,5,6,8}'$
are extracted in five bins of the invariant mass squared of the lepton system, $q^2$. 
We compare our results  for  $P_{4,5,6,8}'$ with Standard Model predictions  including the $q^2$ region in which the LHCb collaboration reported the so-called $P_5'$ anomaly.
\end{abstract}


\maketitle

\tighten

\clearpage


\section{Introduction}

Rare decays of \bms are an ideal probe to search  beyond the Standard Model (SM) of particle physics, since  contributions from new particles lead to effects that are of similar size as the  SM predictions.
The rare decay \bkllzero 
involves the quark transition $b\to s \ell^+ \ell^-$, a flavor changing neutral current  that is forbidden at tree level in the SM.
Higher order SM processes such as penguin or $W^+W^-$ box diagrams allow for such transitions,  leading to branching fractions of less than one in a million.
Various extensions to the SM predict contributions from new physics, which can interfere with the SM amplitudes and lead to enhanced or suppressed branching fractions or modified angular distributions of the decay products. 

We present an angular analysis, using the decay modes \Bto{511339} and \Bto{511335}, in a data sample recorded with the Belle detector. 
The LHCb collaboration reported a discrepancy in the angular distribution of the decay \Bto{511339}, corresponding to a  $3.4\sigma$ deviation from the SM prediction \cite{lhcb2}.
In contrast to the LHCb measurement here the di-electron  channel is also used in this analysis. 

\section{Detector and Datasets}
We use the full $\Upsilon(4S)$ data sample containing $772\times 10^{6}$ $B\bar B$ pairs recorded with the Belle detector \cite{BelleDetektor} at the asymmetric-energy $e^+e^-$ collider KEKB \cite{kekb}.

The Belle detector is a large-solid-angle magnetic
spectrometer that consists of a silicon vertex detector (SVD),
a 50-layer central drift chamber (CDC), an array of
aerogel threshold Cherenkov counters (ACC),  
a barrel-like arrangement of time-of-flight
scintillation counters (TOF), and an electromagnetic calorimeter
comprised of CsI(Tl) crystals (ECL) located inside 
a super-conducting solenoid coil that provides a 1.5~T
magnetic field.  An iron flux-return located outside of
the coil is instrumented to detect $K_L^0$ mesons and to identify
muons (KLM).  The detector
is described in detail elsewhere~\cite{BelleDetektor}.

This analysis is validated and optimized on simulated  Monte Carlo (MC) data. 
The software packages EvtGen \cite{evtgen} and PYTHIA \cite{pythia} are used to simulate the particle decays.
The decay chain is generated, meaning that all intermediate and final state particles are determined.
Final state radiation is calculated by the PHOTOS package \cite{photos}.
The  detector response is simulated with the GEANT3 software package  \cite{geant}.

\section{Reconstruction}
For all charged tracks loose impact parameter constraints are applied with respect to the nominal interaction point in the radial direction  ($|dr| <1.0~\textrm{cm}$) and along the beam direction  ($|dz| <5.0~\textrm{cm}$).
Belle provides a particle identification (PID) likelihood calculated from the energy loss  in the CDC ($\mathrm{d}E/\mathrm{d}x$),  time-of-flight, response of ACC, shape and size of the showers in the ECL and information about hits in the KLM.
Electrons are identified using the likelihood ratio ${\cal P}_{\rm eid}(e)=L(e)/(L(e)+L(\textrm{hadron}))$.
All charged tracks satisfying ${\cal P}_{\rm eid}(e) > 0.1$ are accepted as electrons.
To recover the original momentum of the electrons, a search for photons in a cone of $0.05$ radians around the initial momentum direction of the track is performed.
If photons are found in this region, their momenta are added to the electron.
Charged tracks are accepted as muons if they satisfy the muon likelihood ratio requirement ${\cal P}_{\rm muid}(\mu) > 0.1$.
Charged kaons are selected with the requirement on the likelihood ${\cal P}(K/\pi)=L(K)/(L(K)+L(\pi))>0.1$. 
For  $\pi^\pm$ candidates no PID selection is applied.

$K^\ast$ candidates are formed in the channel $K^{\ast0}\to K^+\pi^-$.
For these candidates, an  invariant mass requirement of $0.6~\textrm{GeV}/c^2 < M_{K^\ast} < 1.4~\textrm{GeV}/c^2$ is applied and a vertex fit is performed, which is used for background suppression later on.

In the final stage of the reconstruction  $K^{\ast}$ candidates are  combined  with oppositely charged lepton pairs to form \bmn candidates. 
The large combinatoric background is suppressed by applying requirements on kinematic variables.
Two independent variables can be constructed using constraints that in \yfs decays  \bms are produced pairwise and each carries half the center--of--mass (CM) frame beam energy, $E_{\textrm{Beam}}$. 
These variables are the beam constrained mass, \mbc, and the energy difference, $\Delta E$, in which signal features a distinct distribution that can discriminate against background.
The variables are defined in the $\Upsilon(4S)$ rest frame as
\begin{align}
	M_\textrm{bc} & \equiv \sqrt{E^2_{\mathrm{Beam}}/c^4 -|\vec p_B|^2/c^2}~\mathrm{and} \\
	\Delta E & \equiv E_B - E_{\mathrm{Beam}},\label{eq:deltaE}
\end{align}
where $E_B$ and $|\vec p_B|$ are the energy and  momentum of the reconstructed candidate, respectively. 
Correctly reconstructed candidates are located around the nominal $B$ mass in \mbc and feature $\Delta E$ of around zero.
Candidates are selected satisfying $ 5.22~ < M_\mathrm{bc}  <~5.3~\mathrm{GeV}/c^2 $ and $ -0.10 \ (-0.05)~ <\Delta E < ~0.05~\mathrm{GeV}$ for $\ell=e$ ($\ell=\mu$).

Large irreducible background contributions  arise from charmonium decays $B\to K^{(\ast)} J/\psi$ and $B\to K^{(\ast)} \psi(2S)$, in which the $c\bar c$ state decays into two leptons. 
These decays have the same signature as the desired signal and are vetoed with the following requirements on $q^2=M_{\ell^+\ell^-}$, the invariant mass of the lepton pair:
\begin{align} \nonumber
	-0.25~\mathrm{GeV}/c^2 &<  M_{ee(\gamma)} - m_{J/\psi}< 0.08~\mathrm{GeV}/c^2, \\ \nonumber
	-0.15~\mathrm{GeV}/c^2 &<  M_{\mu\mu} - m_{J/\psi}< 0.08~\mathrm{GeV}/c^2, \\ \nonumber
	-0.20~\mathrm{GeV}/c^2 &<  M_{ee(\gamma)} - m_{\psi(2S)}< 0.08~\mathrm{GeV}/c^2 ~\mathrm{and} \\ \nonumber
	-0.10~\mathrm{GeV}/c^2 &<  M_{\mu\mu} - m_{\psi(2S)}< 0.08~\mathrm{GeV}/c^2.  \nonumber
\end{align}
In the electron case,  the 4-momentum  of detected photons  from the bremsstrahlung recovery process is added before these requirements are applied. 
Di-electron background can also arise from photon conversions ($\gamma \to e^+ e^-$) and $\pi^0$ Dalitz decays ($\pi^0\to e^+e^-\gamma$).
We require $M_{e(\gamma)e(\gamma)} > 0.14~\textrm{GeV}/c^2$.
For the \bmn candidates, a vertex fit is performed, which is used for background suppression. 
From this fit   the distance between the two leptons along the beam direction $\Delta z_{\ell\ell}$ is also derived.

\section{Background Suppression}


In the selection of $B$ candidates we face   different sources of possible backgrounds.
In continuum background events, $e^+e^-$ annihilates into light quark pairs $u\bar u, d\bar d, s \bar s$ as well as events containing charm quarks $c\bar c$.
These initial quark pairs however exhibit a large energy release, forming back to back jet--like structures.
Combinatorial background arises from incorrect combinations of tracks in $B \bar B$ decays, which is the  dominant source of background.
Finally, a process is referred to as ``peaking background"  when it mimics the signal shape in \mbc. 
For the peaking background several sources have to be taken into account: (1) irreducible background  from \Bjpsi and \Bpsi events, which passes the $q^2$ vetoes; (2) doubly misidentified events  from $B\to K^\ast \pi \pi$ can occur when both pions are misidentified as muons.

\begin{figure*}
	\centering
			\subfigure[$\mathit{NB}_{\rm out}$ for \Bto{511335}]{
				\includegraphics[width=\factorh\textwidth]{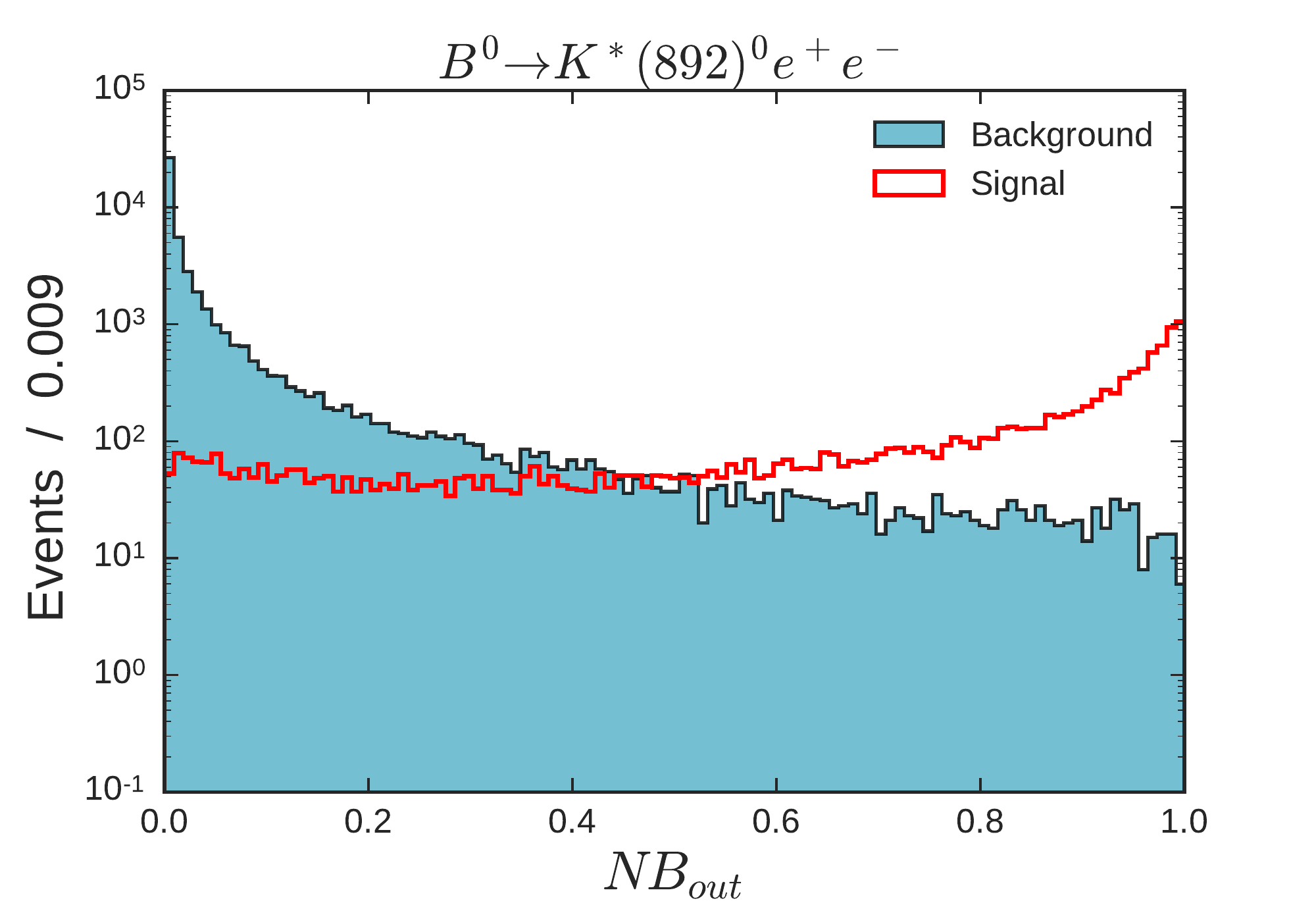}             
			}
			\subfigure[$\mathit{NB}_{\rm out}$ for \Bto{511339}]{
				\includegraphics[width=\factorh\textwidth]{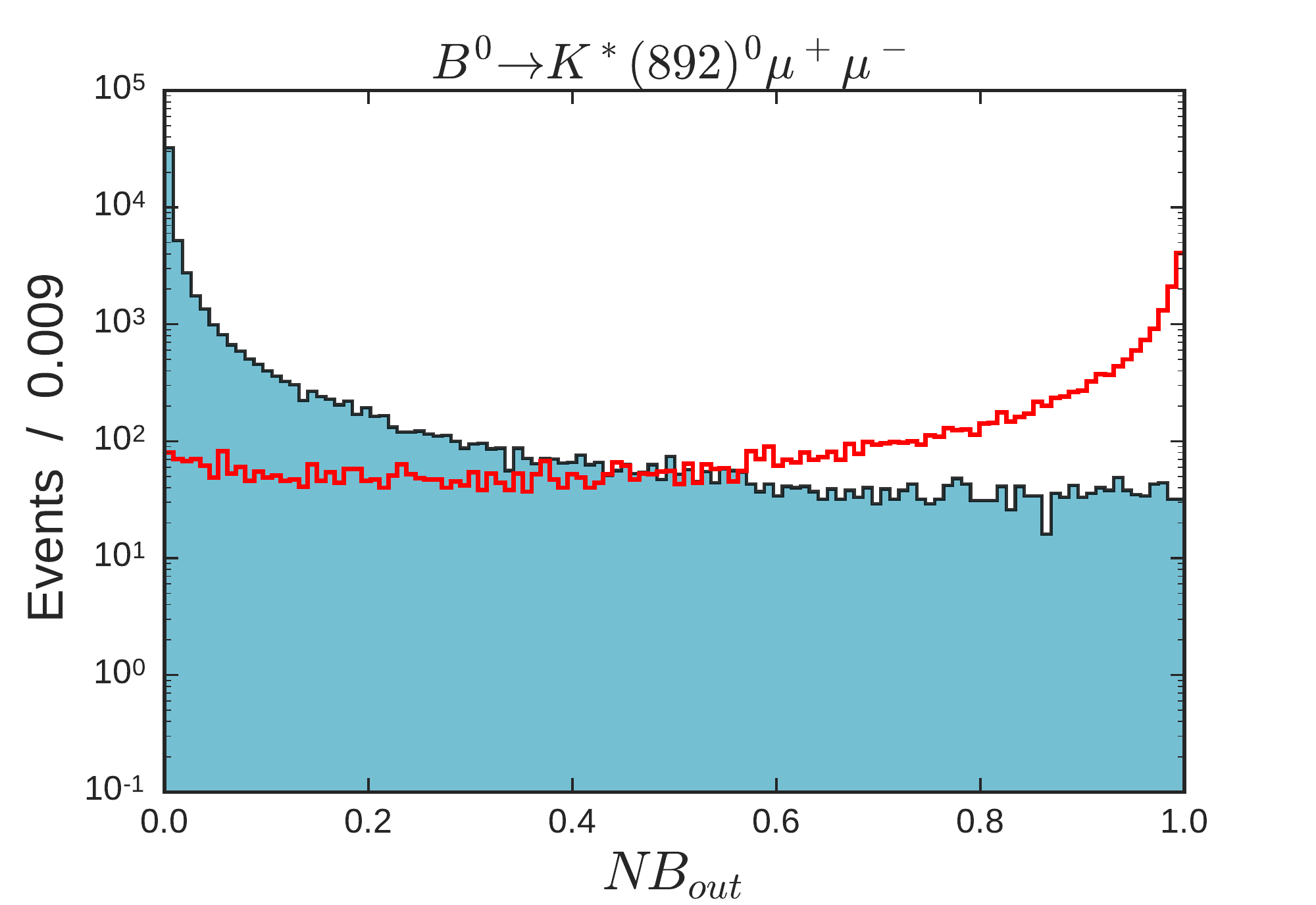}          				       
			}%
	\caption{Performance of the neural networks for the classification of \bkllzero. 
	Signal MC (red line) and simulated background processes from $e^+e^- \to q\bar q \;(q=u,d,s,c,b)$ decays (blue filled) corresponding to  two times the expected size in the Belle dataset are shown.}
	\label{fig:eff_s3}
\end{figure*}

To maximize signal efficiency and purity, neural networks are developed sequentially from the bottom to the top of the decay chain, transferring each time the output probability to the subsequent step so that the most effective selection requirements are applied in the last stage based on all information combined.
All particle candidates are analyzed with a neural network (NeuroBayes \cite{neurobayes})  and an  output, $\mathit{NB}_{\rm out}$,  is assigned.
This output is chosen to correspond to a Bayesian probability in the range $[0,1]$ where the value of one corresponds to signal.
To transfer quality information about the primary particles in the detector to higher  stage composite particles ($K^\ast$ and $B$) the network output of the secondary  particles of each candidate is included as  neural network input. 
In this manner the classifiers for the \bms have $\mathit{NB}_{\rm out}$ for both leptons and the $K^\ast$ included as input.
The classifiers for $e^\pm, \mu^\pm, K^\pm$ and $\pi^\pm$ are taken from the  neural network based full reconstruction, widely used at Belle \cite{fullrecon}.
They use kinematic variables as inputs as well as variables derived from the particle identification system, for instance TOF and KLM information and energy loss in the CDC.
For  $K^\ast$ selection a classifier is trained on simulated data  using kinematic variables and vertex fit information.
The final classification is performed with a requirement on the neural network output $\mathit{NB}_{\rm out}$ for the \bms.
Separate classifiers are trained for \Bto{511339} and \Bto{511335} using event shape variables (i.e. Fox Wolfram Moments \cite{ksfwm}), vertex fit information and kinematic variables.
The most important variables for the neural networks are $\Delta E$, the reconstructed mass of the $K^\ast$, the product of the network outputs of all secondary particles and the distance between the two leptons along the beam direction $\Delta z_{\ell\ell}$.
In case of multiple candidates per event the most probable candidate is chosen, 
based on the neural network output $\mathit{NB}_{\rm out}$.
The final neural network output for signal and background events is displayed in Figure \ref{fig:eff_s3}.

The selection requirement for   the neural networks are optimized for the sensitivity of the angular analysis using pseudo experiments with simulated data, described in Section \ref{sec:angular}.

\section{Signal Yields}

\begin{figure*}
\centering
        \subfigure[\Bto{511339}]{
                \includegraphics[width=\factorh\textwidth]{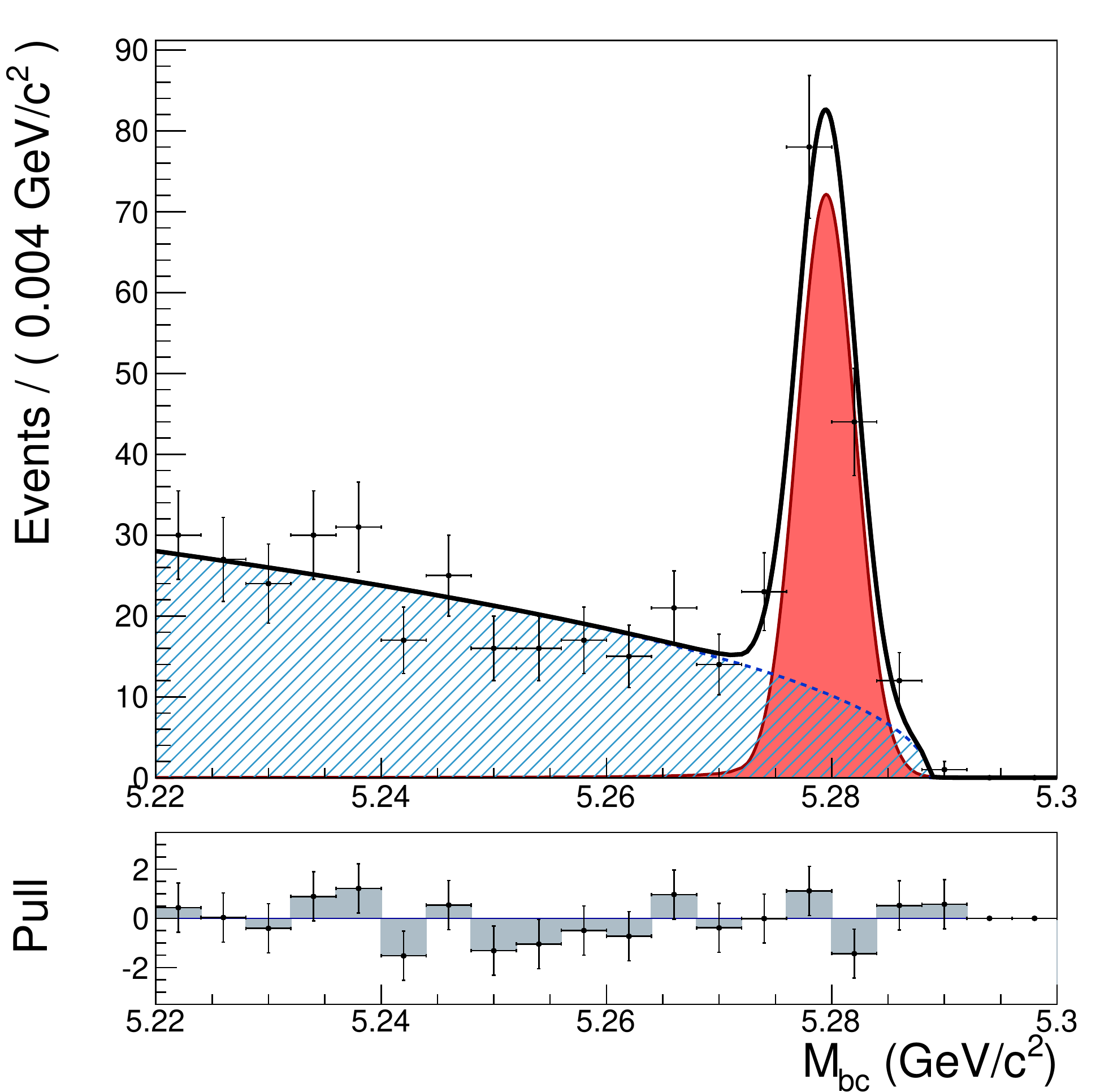}                                 
        }
        \subfigure[\Bto{511335}]{
                \includegraphics[width=\factorh\textwidth]{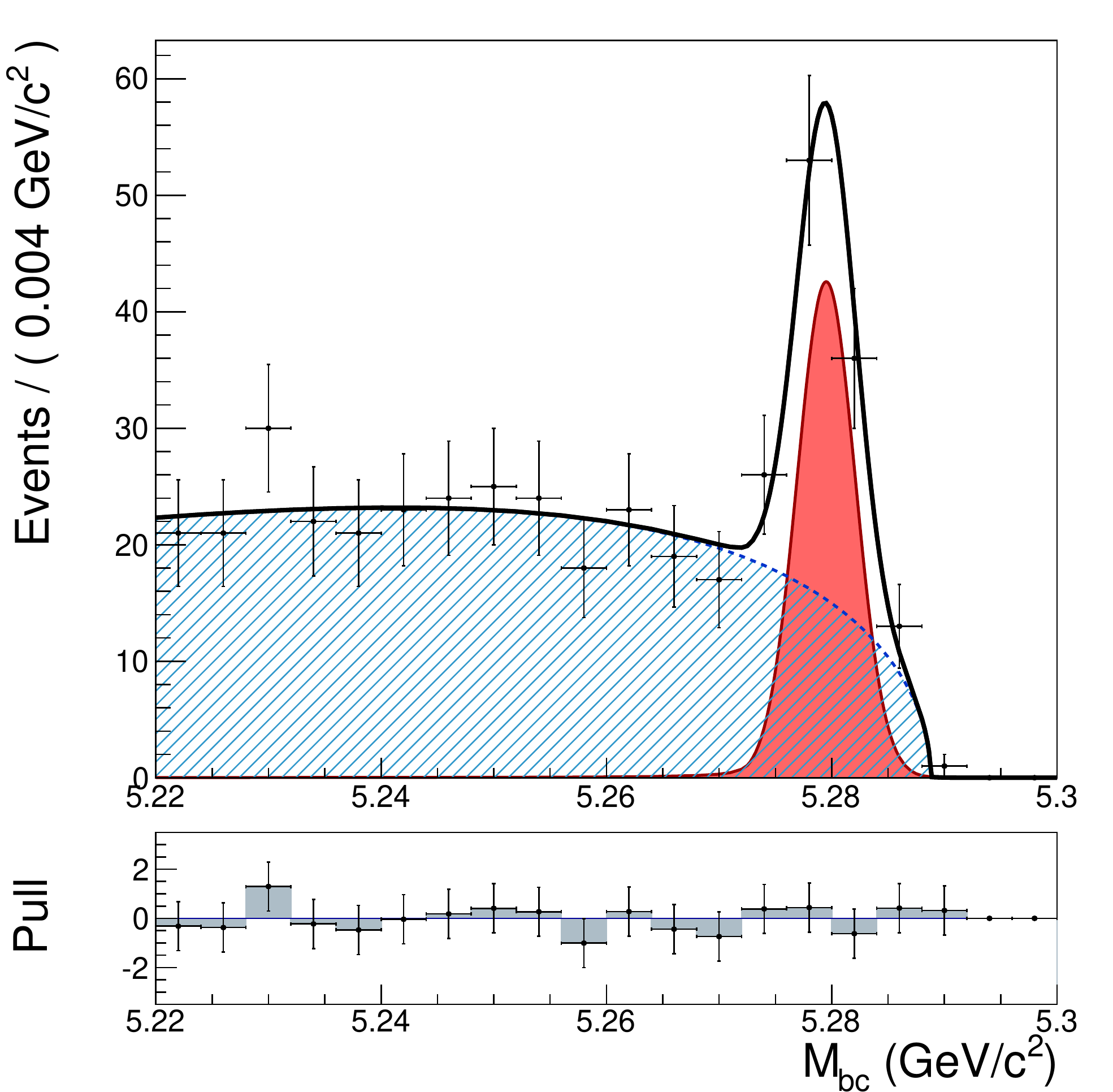}                               
        }%
    \caption{Signal extraction for \bkll on the total range of $q^2$. \fitcomponents}
    \label{fig:mbc_total}
\end{figure*}

Signal and background yields  are extracted  by an unbinned extended maximum likelihood fit to the  \mbc  distribution of  \bkllzero candidates. 
The signal distribution  is parametrized by an empirically determined function introduced by the Crystal Ball Collaboration  \cite{CBshape}.
This so-called Crystal Ball function accounts for radiative tails in the distribution and for  the calorimeter resolution.
All  shape parameters are  determined  by a fit to  data in the control channel $B\to K^\ast J/\psi$ in the corresponding $q^2$ veto region and fixed in the extraction of the \bkllzero yield.
The  background distribution is parametrized by an empirically determined shape introduced by the ARGUS Collaboration \cite{ARGUSshape} and its shape parameters are floated in the fit.
The result of the fits are shown in Figure \ref{fig:mbc_total}.
In the total $q^2$ range there are $118\pm12$ signal candidates  for \Bto{511339} and $69\pm12$ for \Bto{511335}.
For the angular analysis the number of signal events $n_\mathrm{sig}$ and background events $n_{\mathrm{bkg}}$ in the signal region $\mbc >5.27~\mathrm{GeV}/c^2$ are obtained by a fit to \mbc in bins of $q^2$.
The extracted yields and the definition of the bin ranges are presented in   Table \ref{tab:signal_yields}.
As a cross-check, the branching fractions for both modes are calculated and found to be consistent with PDG values within statistical errors.
\begin{table}[htb]
	\centering
	\caption{Fitted yields and statistical error for signal ($n_{\rm sig}$) and background ($n_{\rm bkg}$) events in the binning of $q^2$ for both the combined electron and muon channel.}
	\label{tab:signal_yields}
	\begin{tabular}
		{ @{\hspace{0.5cm}}c @{\hspace{0.5cm}} c
			@{\hspace{0.5cm}}c@{\hspace{0.5cm}}  c@{\hspace{0.5cm}}
		}
		\hline \hline
		 %
		 	Bin &	$q^2$ range in $\textrm{GeV}^2/c^4$ &  $n_\textrm{sig}$ &  $n_\textrm{bkg}$  \\ 
		 		\hline
0 & $1.00 - 6.00$  &  $49.5\pm 8.4$  &  $30.3\pm 5.5$  \\
\hline 
1 &$0.10 - 4.00$  &  $30.9\pm 7.4$  &  $26.4\pm 5.1$  \\
2 &$4.00 - 8.00$  &  $49.8\pm 9.3$  &  $35.6\pm 6.0$ \\
3&$10.09 - 12.90$  &  $39.6\pm 8.0$  &  $19.3\pm 4.4$  \\
4 &$14.18 - 19.00$  &  $56.5\pm 8.7$  &  $16.0\pm 4.0$   \\
\hline \hline
	\end{tabular}
\end{table}

\section{Angular Analysis}
\label{sec:angular}
We perform an angular analysis of \bkllzero\  including the electron and muon modes.
The decay is kinematically    described  by three angles $\theta_\ell$, $\theta_K$ and $\phi$ and the invariant mass squared of the lepton pair $q^2$.
The angle $\theta_\ell$ is defined as the angle between the direction of $\ell^+ ~(\ell^-)$ and the opposite direction of $B~(\bar B)$ in the rest frame of the dilepton system.
The angle $\theta_K$ is defined between the direction of the kaon and the opposite direction of  $B~(\bar B)$ in the $K^\ast$ rest frame.
Finally, the angle $\phi$ is determined as the angle between the decay plane formed by the $\ell^+\ell^-$ system  and the $K^\ast$ decay plane.
Definitions of the angles follow Ref. \cite{lhcbafb}.
The analysis is performed in four bins of $q^2$ with an additional zeroth bin in the range	$1.0 < q^2 < 6.0 ~\mathrm{GeV}^2/c^4$, which is considered to be the cleanest regarding form-factor uncertainties \cite{Altmannshofer:2008dz}.
The binning in $q^2$ is detailed in Table \ref{tab:signal_yields} together with the measured signal and background yields.
Uncovered regions in the $q^2$ spectrum arise from vetoes against backgrounds of the charmonium resonances $J/\psi \to \ell^+\ell^-$ and $\psi(2S) \to \ell^+\ell^-$ and vetos against $\pi^0$ Dalitz decays and photon conversion.


The full angular distribution of $B \to K^{\ast 0}(\to K^\pm \pi^\mp) \ell^+ \ell^-$ can be parametrized using  definitions presented in Ref. \cite{Altmannshofer:2008dz} by
\begin{widetext}
\begin{align}
\frac{1}{\textrm{d}\Gamma/\textrm{d} q^2}
\frac{\textrm{d}^4\Gamma}{\textrm{d}\cos\theta_\ell\; \textrm{d}\cos\theta_K\; \textrm{d}\phi\; \textrm{d}q^2}
= 
&\frac{9}{32\pi}
\left[
\frac{3}{4} (1 - F_L) \sin^2\theta_K 
+F_L\cos^2\theta_K \right.\nonumber \\
&+\frac{1}{4}(1-F_L)\sin^2\theta_K\cos2\theta_\ell \nonumber \\
&-F_L \cos^2\theta_K\cos 2\theta_\ell + S_3 \sin^2\theta_K \sin^2\theta_\ell\cos 2\phi  \nonumber  \\
& + S_4 \sin 2\theta_K \sin 2\theta_\ell \cos\phi + S_5 \sin 2\theta_K\sin\theta_\ell\cos\phi \nonumber \\
&+S_6 \sin^2\theta_K\cos\theta_\ell + S_7 \sin 2\theta_K \sin\theta_\ell\sin\phi \nonumber \\ 
&+  S_8 \sin 2\theta_K\sin 2\theta_\ell\sin\phi + S_9 \sin^2\theta_K\sin^2\theta_\ell\sin 2\phi \;\bigg],\label{eq:signalpdf}
\end{align}
\end{widetext}
where the  observables $F_L$ and $S_i$ are functions of $q^2$ only.
The observables are functions of Wilson coefficients, containing information about the short-distance effects and can be affected by new physics. 
The observables $P_i'$, introduced in Ref. \cite{DescotesGenon:2012zf}, defined as
\begin{equation}
P'_{i=4,5,6,8} = \frac{S_{j=4,5,7,8}}{\sqrt{F_L(1-F_L)}},
\end{equation}
are considered to be largely free from form-factor uncertainties \cite{p_theory}.
In total, there are eight free parameters, which can be obtained from a  fit  to the data.
The   statistics in this analysis  are not sufficient to perform an eight-dimensional fit.
In the following a folding technique is described, which reduces the number of fitting parameters and hence improves the convergence of the fit.
The folding is applied to specific regions in the three-dimensional angular space, exploiting the symmetries of the cosine and sine functions to cancel  terms in Eq. (3).
As a consequence the number of free parameters in the fit is reduced  without losing experimental sensitivity.
This procedure is explained in more detail in Refs.~\cite{lhcb1} and \cite{lhcb_phd}.
With the following transformations to the dataset one can be  sensitive to the observable of interest:
\begin{equation}
P_4',S_4:\;\begin{cases}
\phi \to -\phi & \text{for $\phi<0$}\\
\phi \to \pi - \phi & \text{for $\theta_\ell>\pi/2$}\\
\theta_\ell \to \pi -\theta_\ell & \text{for $\theta_\ell>\pi/2$},
\end{cases}
\label{eq:foldingp4}
\end{equation}
\begin{equation}
P_5',S_5:\;\begin{cases}
\phi \to -\phi & \text{for $\phi<0$}\\
\theta_\ell \to \pi -\theta_\ell & \text{for $\theta_\ell>\pi/2$},
\end{cases}
\label{eq:foldingp5}
\end{equation}
\begin{equation}
P_6',S_7:\;\begin{cases}
\phi \to \pi -\phi & \text{for $\phi>\pi/2$}\\
\phi \to - \pi - \phi & \text{for $\phi < -\pi/2$}\\
\theta_\ell \to \pi -\theta_\ell & \text{for $\theta_\ell>\pi/2$},
\end{cases}
\label{foldingp6}
\end{equation}
\begin{equation}
P_8',S_8:\;\begin{cases}
\phi \to \pi -\phi & \text{for $\phi>\pi/2$}\\
\phi \to - \pi - \phi & \text{for $\phi < -\pi/2$}\\
\theta_K \to  \pi - \theta_K & \text{for $\theta_\ell > \pi/2$}\\
\theta_\ell \to \pi -\theta_\ell & \text{for $\theta_\ell>\pi/2$}.
\end{cases}
\label{eq:foldingp8}
\end{equation}
Each of the transformations causes all $S_i$ terms of Eq. (3), except for $S_3$ and the corresponding $S_i$ term, to vanish.
The number of free parameters of each transformed decay rate is consequently reduced to three: $F_L$, $S_3$ and one of the observables $S_{4,5,7,8}$ or  $P'_{4,5,6,8}$.

One can extract the transverse polarization asymmetry $A^{(2)}_T$   with the transformation:  $A^{(2)}_T = 2S_3/(1-F_L)$.
To parameterize the background we use smoothed template histograms.
A three-dimensional PDF is constructed by multiplying the histograms of each projection of the angular variables:
 \begin{align}
 f^\textrm{hist}_\textrm{bkg}(q^2, \cos\theta_\ell, \cos\theta_K, \phi) \nonumber
 = \\ \nonumber 
 h_1(q^2,\cos\theta_\ell)   \cdot 
 h_2(q^2,\cos\theta_K) \cdot \nonumber
 h_3(q^2,\phi).\nonumber
 \end{align}
 This method is fast and robust even if the background shape is complex.
 The correlation in the background sample between the observables is negligible allowing for this procedure.
 However, it introduces systematic deviations from the true PDF due to statistical fluctuations.
 To compensate for this, the histograms are smoothed with an algorithm introduced in Ref.~\cite{blobel}, which takes into account Poisson errors for bins with a small number of entries.
 This method aims to optimize the pull distribution from smoothed histograms to the original histogram with statistical fluctuations by a least square minimization.

All methods are tested in toy MC studies using simulated events where each measurement is performed 10,000 times.
 The most important optimization is that of the neural network requirement for both \Bto{511335} and \Bto{511339}, which determines the signal to background ratio in the fit.
The sensitivity is optimized by minimizing the total error of $P_5'$ in $q^2$ bin 2.
In the toy studies this is calculated as the linear sum of the mean statistical error from the toy study $e_{stat}$, the systematic error from a fit bias $e_{bias}$ and the estimated error from peaking background $\hat{e}_{peaking}$.
From this procedure, the estimated sensitivity for each pair of requirements is obtained.
%

\section{Acceptance and Efficiency}

\begin{figure*}
	\centering
	\subfigure[Efficiency in $\phi$]{
		\includegraphics[width=\factorh\textwidth]{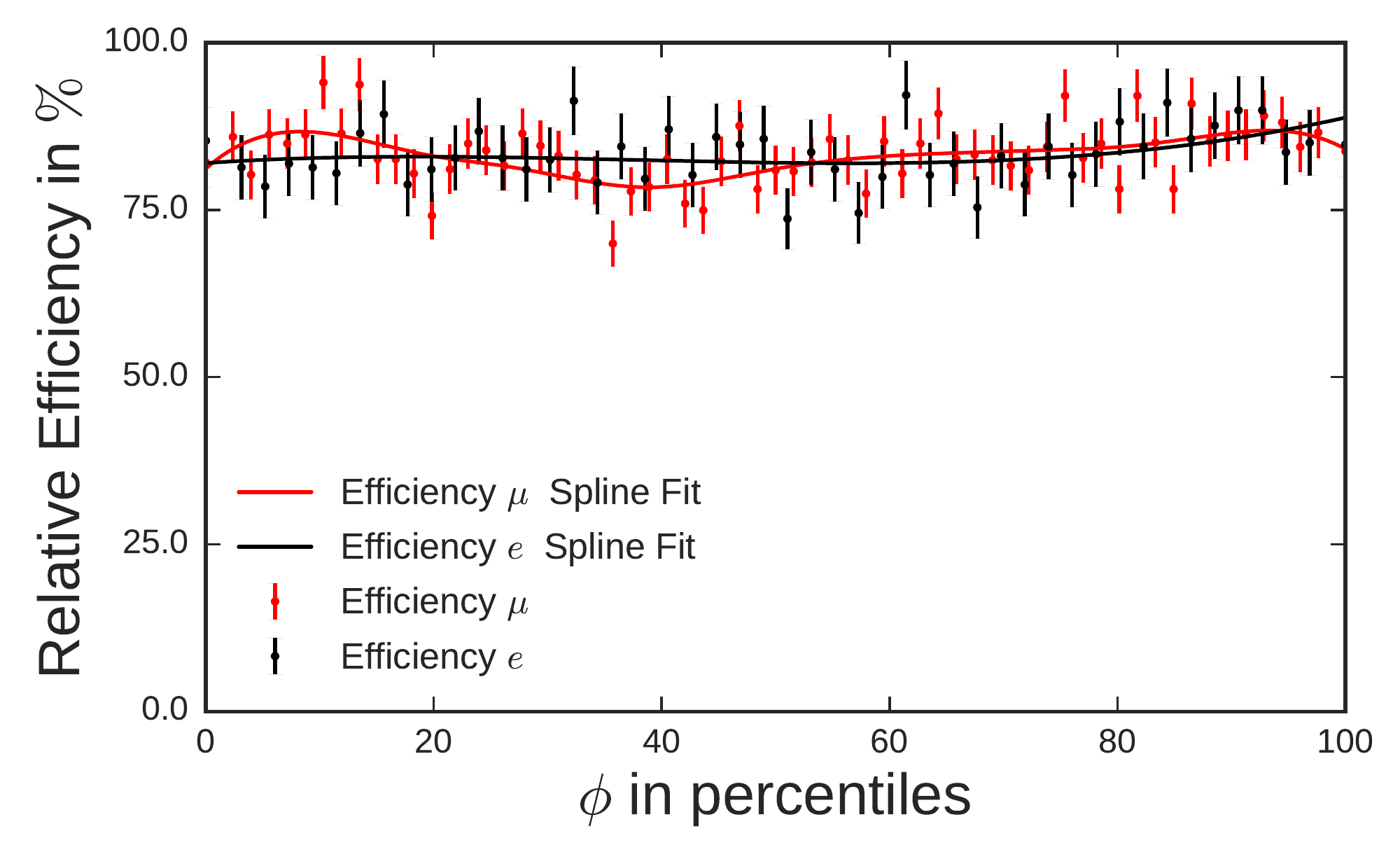}          		    
	}%
	\subfigure[Efficiency in $\theta_\ell$]{
		\includegraphics[width=\factorh\textwidth]{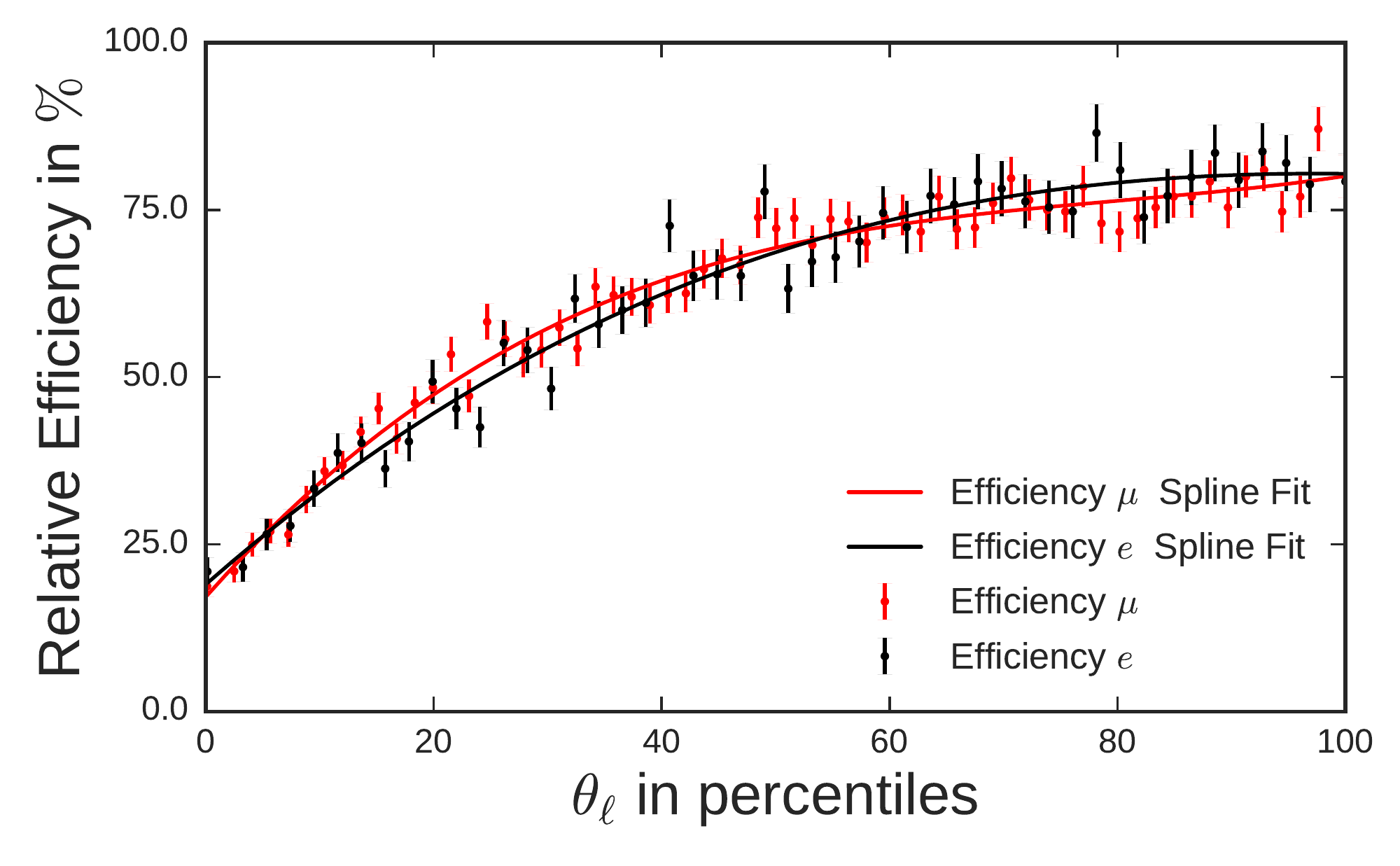}          		       
	}\\
	\subfigure[Efficiency in $\theta_K$]{
		\includegraphics[width=\factorh\textwidth]{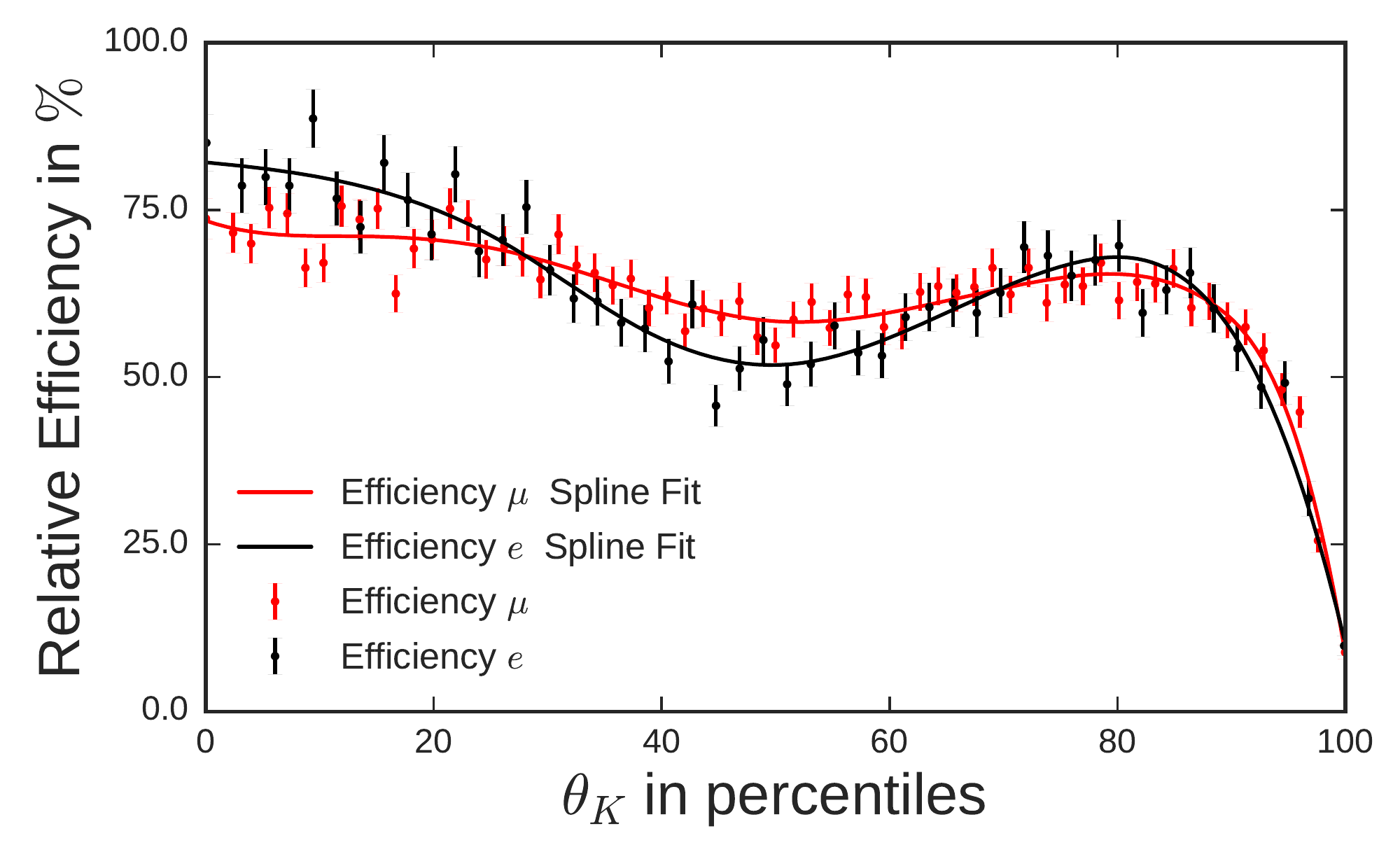}   
	}%
	\subfigure[Efficiency in $q^2$]{
		\includegraphics[width=\factorh\textwidth]{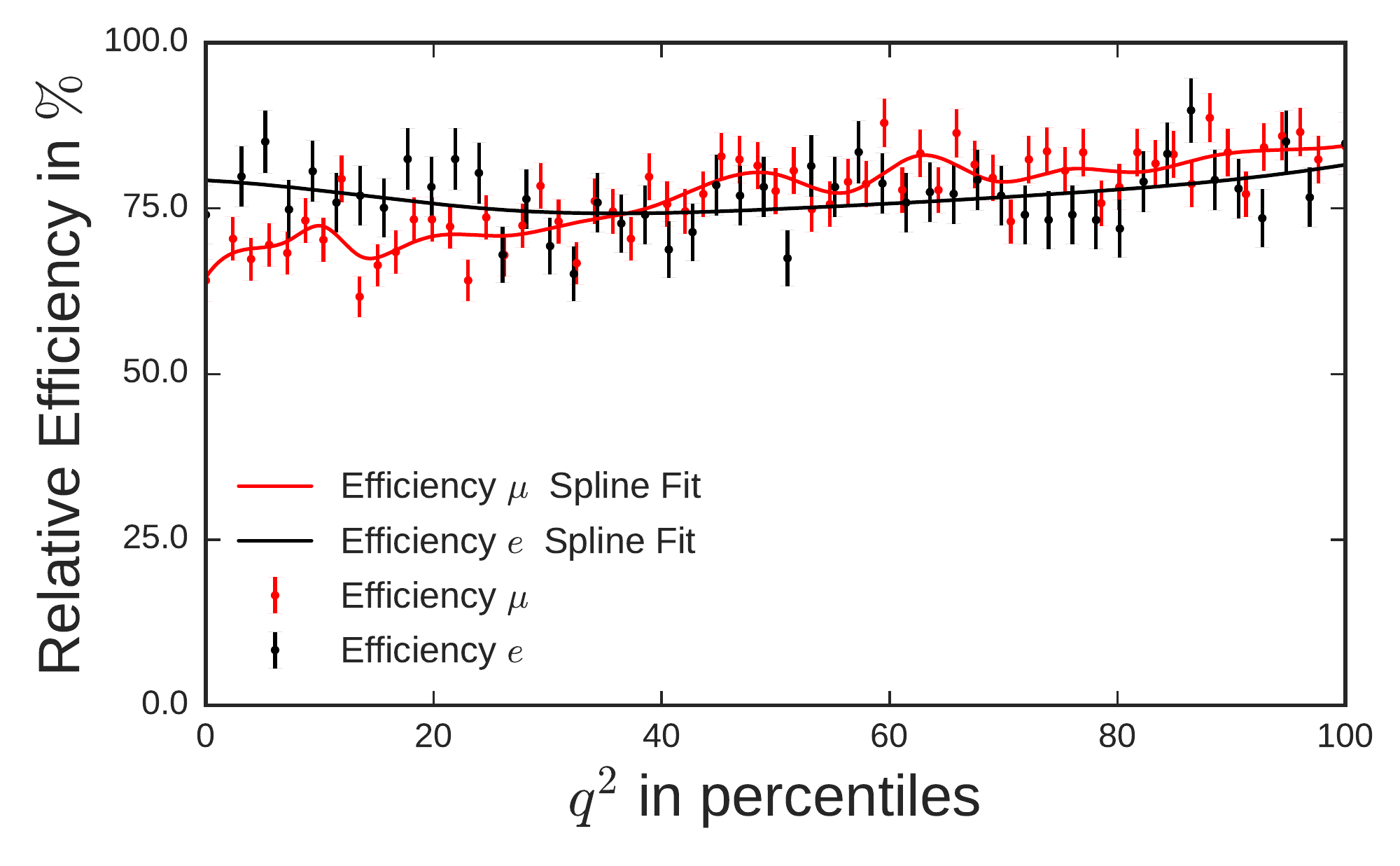}         
	}%
	\caption{Example for the fits of the efficiencies in $q^2$ bin 2 . Both the efficiency fits  (solid lines) for the electron (black) and muon (red) channel are superimposed over the ratio of generated and reconstructed events (data points).}
	\label{fig:eff_statistical_binning}
\end{figure*}

We account for acceptance and efficiency effects in the fit by assigning weights to the data. 
We weight each event by the inverse of its combined efficiency, which is derived from the direct product of the  efficiencies of the  angular observables and  $q^2$.
The individual reconstruction efficiency  for each observable $x$ is obtained by extracting the differences between the reconstructed and generated distributions.
In order to minimize statistical fluctuations in this process, the generated distribution is transformed to a flat distribution.
For $x_{\rm gen}$ from generated events in  the signal MC the corresponding cumulative density distribution is  derived by a spline interpolation  $s_{\rm gen}$ from a histogram of the cumulative density distribution of $x_{\rm gen}$. 
In the next step we transform the $x_{\rm rec}$ value from reconstructed signal MC events, which include  reconstruction and acceptance effects, with   $s_{\rm gen}$  and derive the distribution of  the  reconstruction efficiency.
The distribution of the  reconstruction efficiency  follows
\begin{equation}
x_{\rm eff} = s_{\rm gen}(x_{\rm rec}).
\end{equation}
The final efficiency for each observable is then  fitted with a spline fit $s_{\rm eff}$ to the distribution of $x_{\rm eff}$, which fits orthogonal splines to the data so that the pull between the fit and the data-points becomes  a Gaussian with width one and mean zero.
Finally, the efficiency $\epsilon(x)$ for observable $x$ is calculated from 
\begin{equation}
\epsilon(x) = s_{\rm eff}(s_{\rm gen}(x)).
\end{equation}
The combined efficiency $f^{\rm bin}_{\rm eff}$ is determined in each bin of $q^2$ and is calculated by
\begin{align}
f^{\rm bin}_{\rm eff}(\cos\theta_\ell,\cos\theta_K,\phi,q^2) = \nonumber
& \epsilon(\cos\theta_\ell)\otimes
\epsilon(\cos\theta_K)\otimes  \epsilon(\phi)\otimes  \linebreak  \\ 
 &\epsilon(q^2),
\label{eq:angular_eff}
\end{align}
assuming that the efficiency is uncorrelated in the three-dimensional angular space, which is validated for the systematic uncertainties.
The fits for the efficiencies in the $q^2$ range $4 < q^2 < 8~\gevsq$ are shown in Figure \ref{fig:eff_statistical_binning}.
In the final fit the weights are normalized, so that the sum of all weights equals the total number of events in the fit.

\section{Fit Procedure}
The signal and background fractions are derived from a fit to \mbc beforehand, where the yields are listed in Table \ref{tab:signal_yields}.
The \mbc variable is split into a signal (upper) and sideband (lower) region at $5.27~\mathrm{GeV}/c^2$. 
In the second step the shape of the background for the angular observables is estimated on the \mbc sideband.
This is possible as the angular observables have shown to be uncorrelated to \mbc in the background sample.

All observables $P_{4,5,6,8}'$ are extracted  from the data in the signal region using three-dimensional unbinned maximum likelihood fits in four bins of $q^2$ and the additional zeroth bin  using the folded signal PDF, fixed background shapes and a fixed number of signal events.
Each $P_{4,5,6,8}'$ is fitted  with $F_L$ and $A_T^{(2)}$.
Counting also the zeroth bin, which exhibits overlap with the  range of the first and second bin,   20  measurements are performed.

\section{Systematic Studies}

For the angular analysis  sources of systematic uncertainty are considered, if they introduce an angular or $q^2$ dependent bias to the distributions of signal or background candidates.
Systematic uncertainties  are examined using pseudo-experiments with large signal yields in order to minimize statistical fluctuations and compare the nominal  with a varied model.
The  variation between the average of two models is taken as systematic uncertainty.
Observed differences between data and MC are modeled within the fit for the efficiency correction as a bias.
A systematic error is derived from the difference between the results from a fit with the nominal efficiency correction and the modified correction including differences observed on data.
Due to the limited  number of candidates in some measurements, a fit bias is observed in some bins of the angular analysis.
In 10,000 pseudo experiments on simulated data the fit for each measurement is performed and the results are compared to the simulated values.
The mean of the pull distribution from the toy study is used for each measurement to determine a systematic bias on the measurement.
The central values of the measurements are not corrected for the bias  but  the absolute value of the deviation is assigned as a systematic error.
For the fit of the reconstruction efficiency function a factorization of the efficiencies in the angular observables and $q^2$ is assumed, which is not the case for \ctl in the low $q^2$ region.
The deviation in a simulated dataset with efficiency correction weights and a  dataset based on generator truth  is evaluated.
The difference  between the two fits is taken into account as a systematic uncertainty for the efficiency correction in the fit.

Peaking backgrounds are estimated  for each $q^2$ bin  using MC.
In total  less than six such background events  are expected  in the muon channel, and less than one in the electron channel.
The impact of the peaking component is simulated by repeating the toy study and replacing six randomly selected events from the signal with events from the peaking background in each bin. 
The mean deviation of the procedure is $\pm0.027$ for the value of $P_{4,5,6,8}'$, which corresponds to approximately $2-5\%$ of the statistical error.
The signal cross-feed is calculated for the $B^0$ decay channels (\Bto{511339},\Bto{511335},\Bto{511336},\Bto{511332}) and found to be insignificant.
The parametrization in Eq. (3) does not include a potential S-wave contribution from $K^\ast(892)$ decays.
The fraction $F_S$ is searched for  in our data by fitting the invariant mass of the $K \pi$ pair resulting in  $F_S$ being consistent with zero with a small uncertainty.
If there is a production, detection or direct  $\cal CP$ asymmetry observed, the measured $\cal CP$-symmetric parameters,  must be corrected.
Since the yields of $B^0$ and $\bar B^0$ events are statistically equal in the signal region of our measurement with 153 and 150 events, respectively, and the  theoretical values are small for the $\cal CP$ asymmetric parameters $A^{(s)}_i$ ($\lesssim {\cal O}(10^{-3})$  \cite{Altmannshofer:2008dz}) influences of this kind are neglected.
All sources of included systematics are summarized separately for $P_4'$, $P_5'$, $P_6'$ and $P_8'$ in Tables \ref{tab:sys_allp4}, \ref{tab:sys_allp5}, \ref{tab:sys_allp6} and \ref{tab:sys_allp8}. 
The total systematic uncertainty is calculated as the square root of the quadratic sum of all systematic uncertainties.

\begin{table}[htb]
	\centering
	\caption{Summary of all  systematic uncertainties for $P'_4$.}
	\label{tab:sys_allp4}

	\begin{tabular}{lccccc}
	\hline\hline
		Bin        &   0    &   1    &   2    &   3    &   4    \\
	\hline

Peaking Background  & 0.0855 & 0.0646 & 0.0366 & 0.0457 & 0.0358 \\ 
Data/MC Difference  & 0.0109 & 0.0088 & 0.0020 & 0.0003 & 0.0047 \\ 
Efficiency Correction  & 0.1475 & 0.0241 & 0.0599 & 0.0877 & 0.0650 \\ 
Fit Bias  & 0.0316 & 0.0114 & 0.0007 & 0.0558 & 0.0027 \\ 
\hline
Total & 0.1738 & 0.0704 & 0.0702 & 0.1135 & 0.0744 \\

	\hline\hline
	\end{tabular} 
\end{table}

\begin{table}[htb]
	\centering
	\caption{Summary of all  systematic uncertainties for $P'_5$.}
	\label{tab:sys_allp5}

	\begin{tabular}{lccccc}
	\hline\hline
		Bin        &   0    &   1    &   2    &   3    &   4    \\
	\hline

Peaking Background  & 0.0901 & 0.0636 & 0.0078 & 0.0498 & 0.0131 \\ 
Data/MC Difference  & 0.0112 & 0.0067 & 0.0208 & 0.0142 & 0.0029 \\ 
Efficiency Correction  & 0.0397 & 0.0205 & 0.0098 & 0.0215 & 0.0327 \\ 
Fit Bias  & 0.0031 & 0.0061 & 0.0430 & 0.0127 & 0.0460 \\ 
\hline
Total & 0.0992 & 0.0675 & 0.0494 & 0.0575 & 0.0580 \\

	\hline\hline
	\end{tabular} 
\end{table}

\begin{table}[htb]
	\centering
	\caption{Summary of all  systematic uncertainties for $P'_6$.}
	\label{tab:sys_allp6}

	\begin{tabular}{lccccc}
	\hline\hline
		Bin        &   0    &   1    &   2    &   3    &   4    \\
	\hline

Peaking Background  & 0.0170 & 0.0513 & 0.0229 & 0.0215 & 0.0026 \\ 
Data/MC Difference  & 0.1298 & 0.1378 & 0.1655 & 0.2201 & 0.2341 \\ 
Efficiency Correction  & 0.0835 & 0.0432 & 0.0683 & 0.0218 & 0.0192 \\ 
Fit Bias  & 0.0735 & 0.1189 & 0.0562 & 0.0027 & 0.0268 \\ 
\hline
Total & 0.1718 & 0.1939 & 0.1890 & 0.2222 & 0.2364 \\

	\hline\hline
	\end{tabular} 
\end{table}

\begin{table}[htb]
	\centering
	\caption{Summary of all  systematic uncertainties for $P'_8$.}
	\label{tab:sys_allp8}

	\begin{tabular}{lccccc}
	\hline\hline
		Bin        &   0    &   1    &   2    &   3    &   4    \\
	\hline

Peaking Background  & 0.1242 & 0.0161 & 0.0395 & 0.0518 & 0.0255 \\ 
Data/MC Difference  & 0.1433 & 0.1630 & 0.1531 & 0.1955 & 0.2316 \\ 
Efficiency Correction  & 0.0319 & 0.0824 & 0.0359 & 0.0418 & 0.0099 \\ 
Fit Bias  & 0.0337 & 0.1033 & 0.0579 & 0.0048 & 0.0047 \\ 
\hline
Total & 0.1952 & 0.2105 & 0.1722 & 0.2065 & 0.2332 \\

	\hline\hline
	\end{tabular} 
\end{table}

\section{Results}
\label{subsec:smtheo}

\begin{figure*}
	\centering
	\subfigure[Result for $P_4'$]{
		\includegraphics[width=\factorh\textwidth]{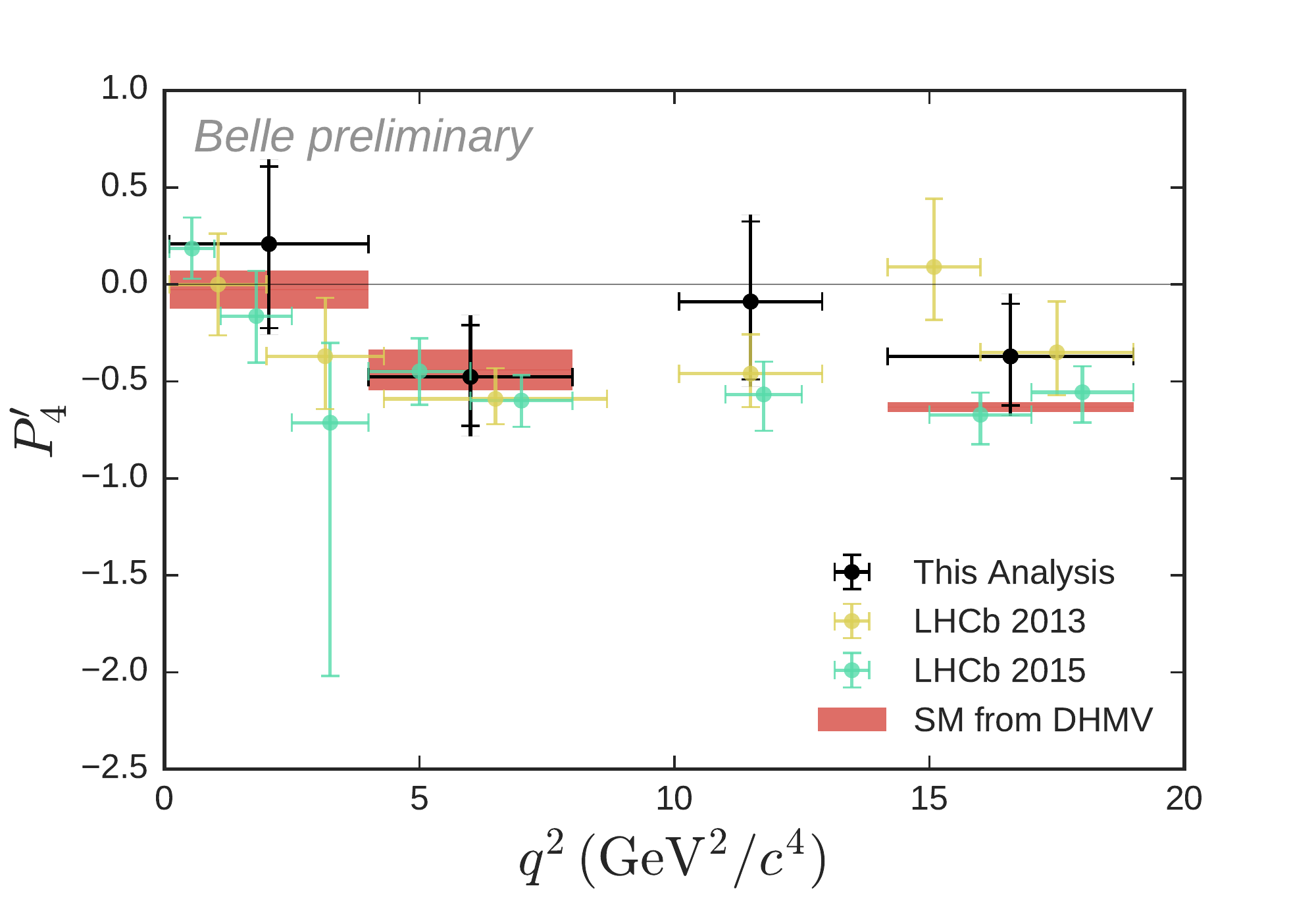}          
	}%
	\subfigure[Result for $P_5'$]{
		\includegraphics[width=\factorh\textwidth]{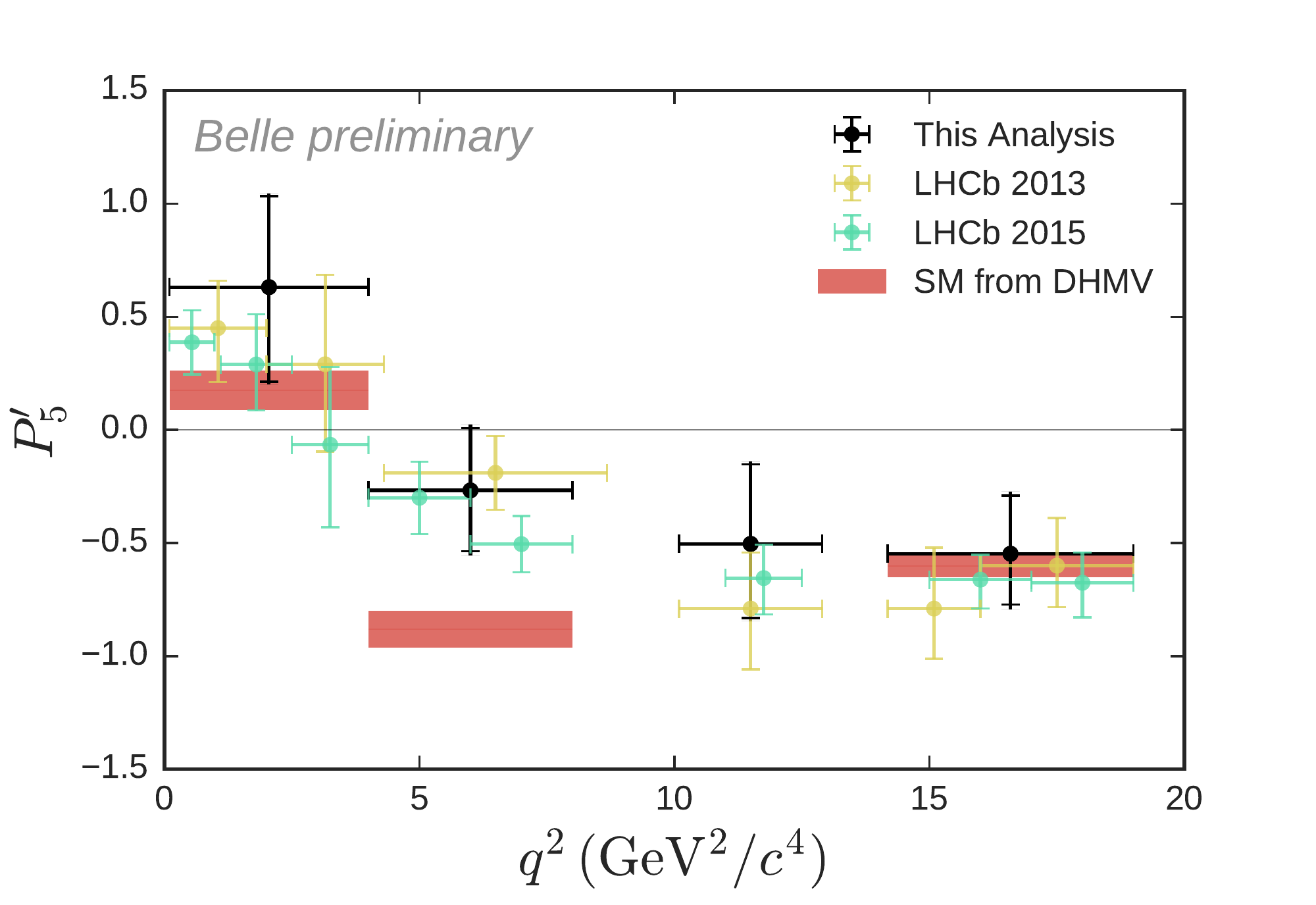}          
	}\\
	\subfigure[Result for $P_6'$]{
		\includegraphics[width=\factorh\textwidth]{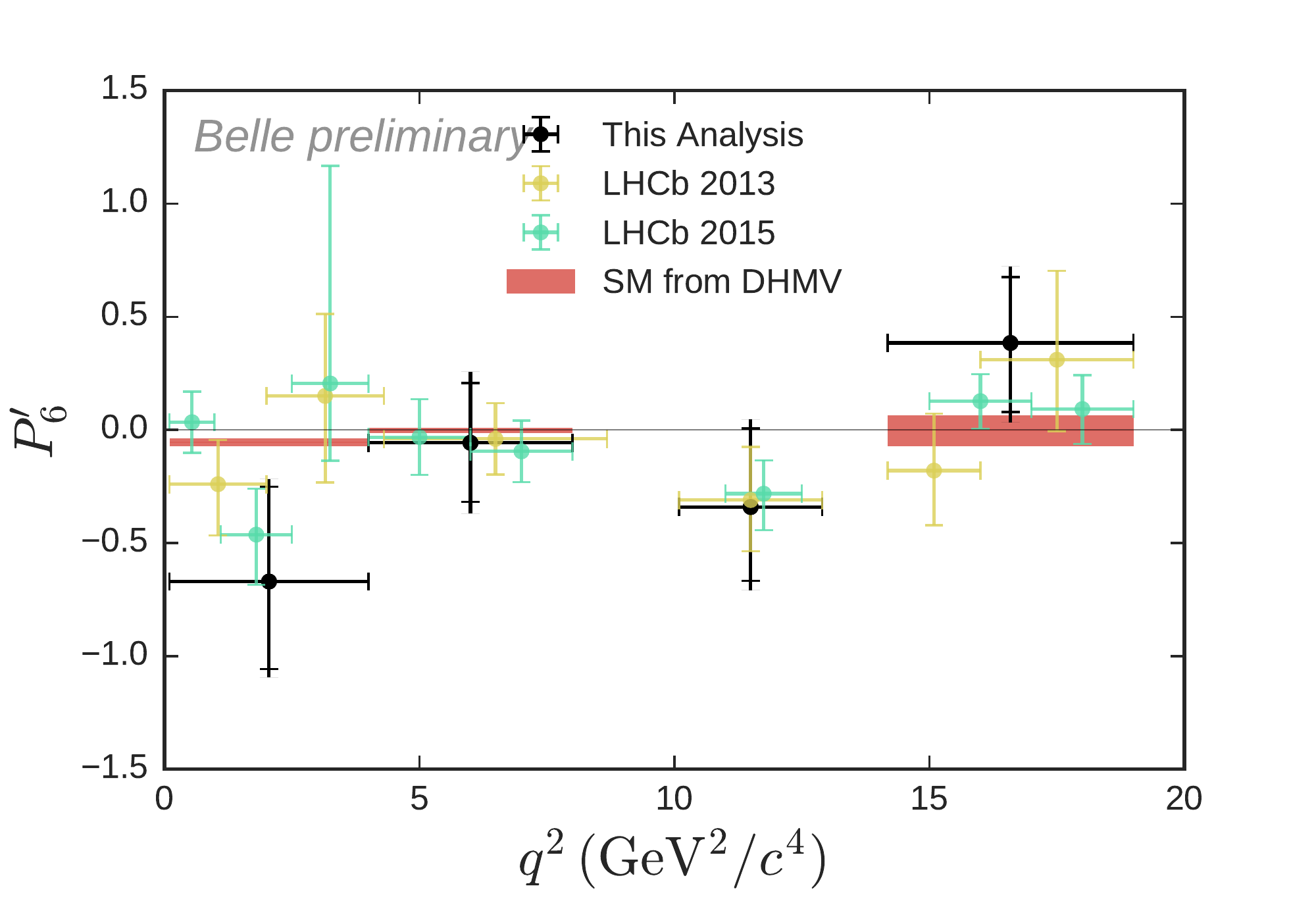}          
	}%
	\subfigure[Result for $P_8'$]{
		\includegraphics[width=\factorh\textwidth]{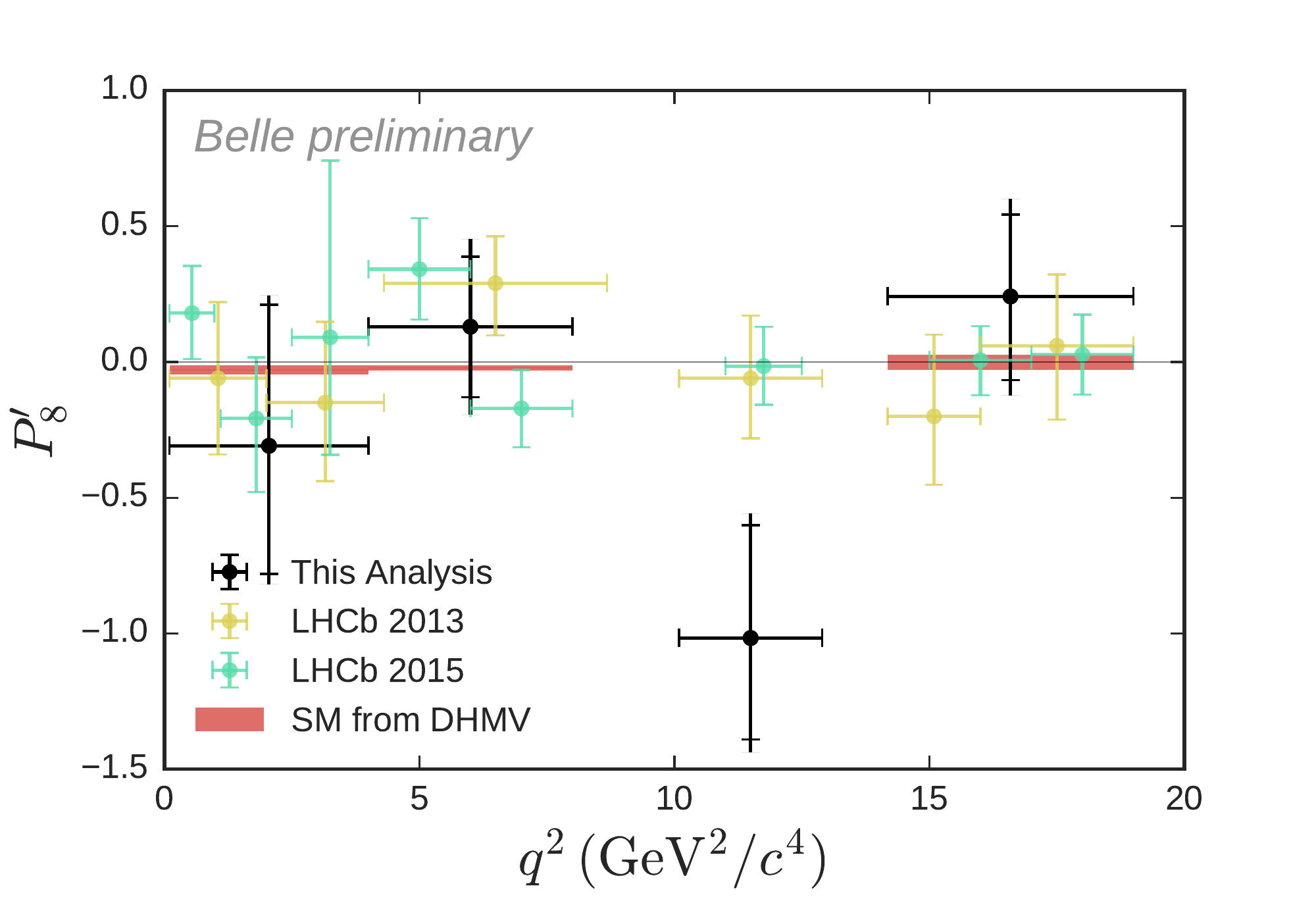}          
	}%
	\caption{Result for the $P'$ observables compared to SM predictions from various sources described in Section \ref{subsec:smtheo}. Results from LHCb \cite{lhcb1,lhcb2} are shown for comparison. }
	\label{fig:res}
\end{figure*}

The measurements are compared with SM predictions based upon different theoretical calculations.
Values from DHMV  refer  to the soft-form-factor method of Ref. \cite{Descotes-Genon:2014uoa}, which is also used in the LHCb measurement.
BSZ corresponds to using  QCD form factors computed from LCSRs with $K^\ast$ distribution amplitudes described in \cite{Straub:2015ica}.
The third set of theoretical predictions is provided  by the methods and authors of Refs. \cite{Jager:2012uw,Jager:2014rwa} whose framework is specially tailored to the low $q^2$ region. 
It is referred to as JC.
The results are listed in Table \ref{tab:angular_results} and  are shown in Figure \ref{fig:res} together with available SM prediction from DHMV  and LHCb measurements.

\begin{figure*}
\centering

    \subfigure[Projection in $\theta_K$]{
                \includegraphics[width=\factort\textwidth]{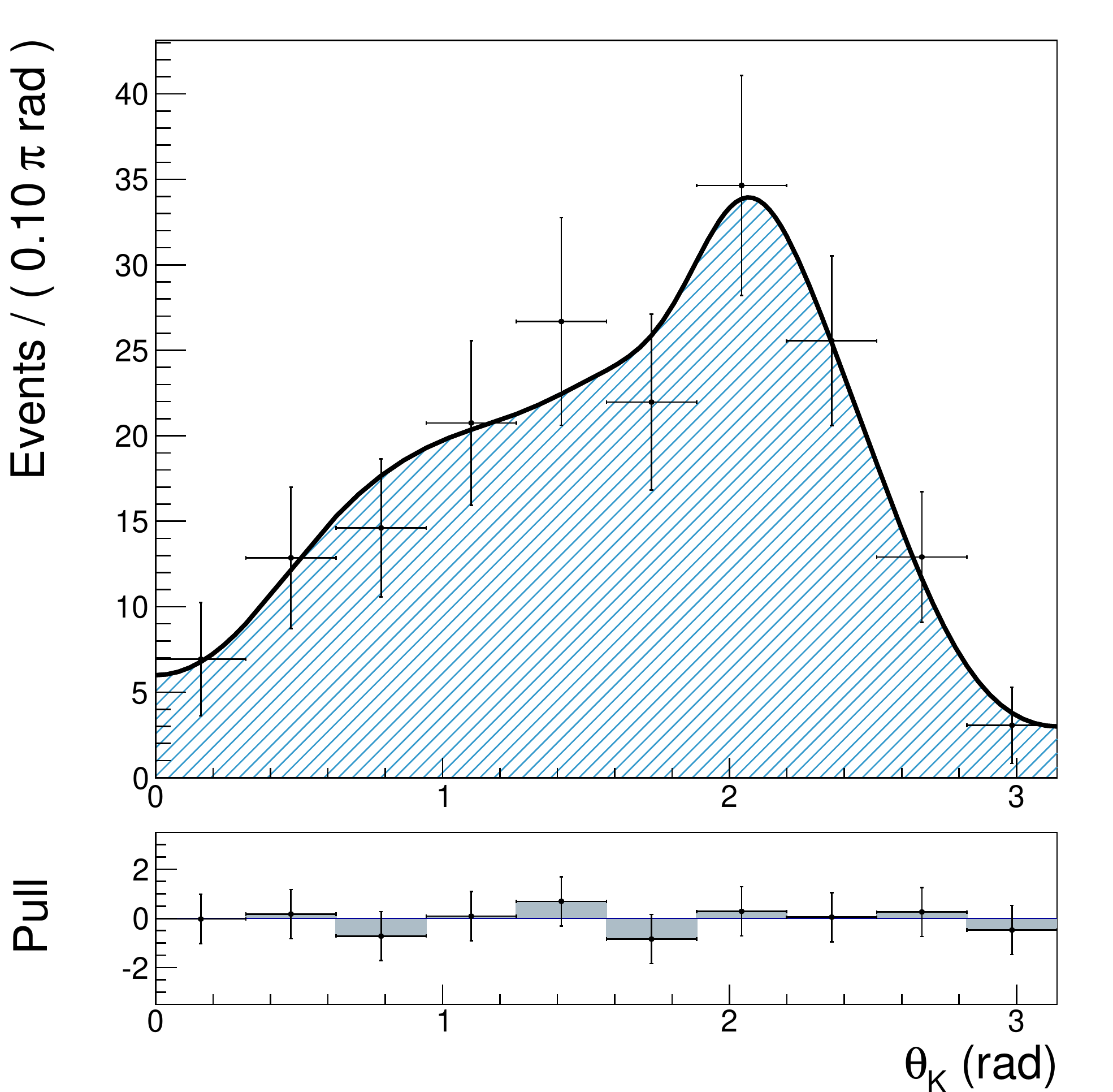}                                 
    }%
    \subfigure[Projection in $\theta_\ell$]{
                \includegraphics[width=\factort\textwidth]{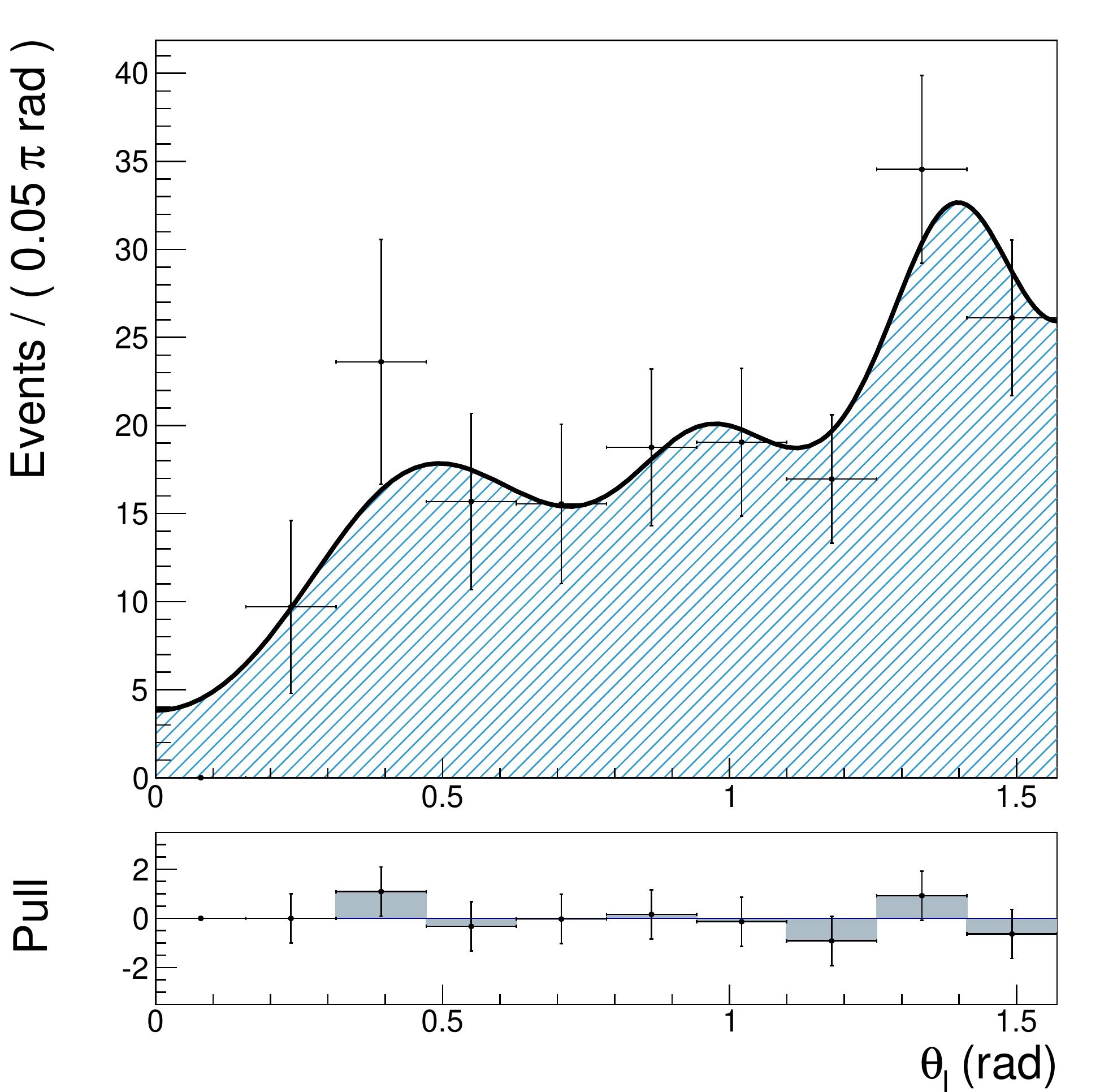}          
    }%
    \subfigure[Projection in $\phi$]{
                \includegraphics[width=\factort\textwidth]{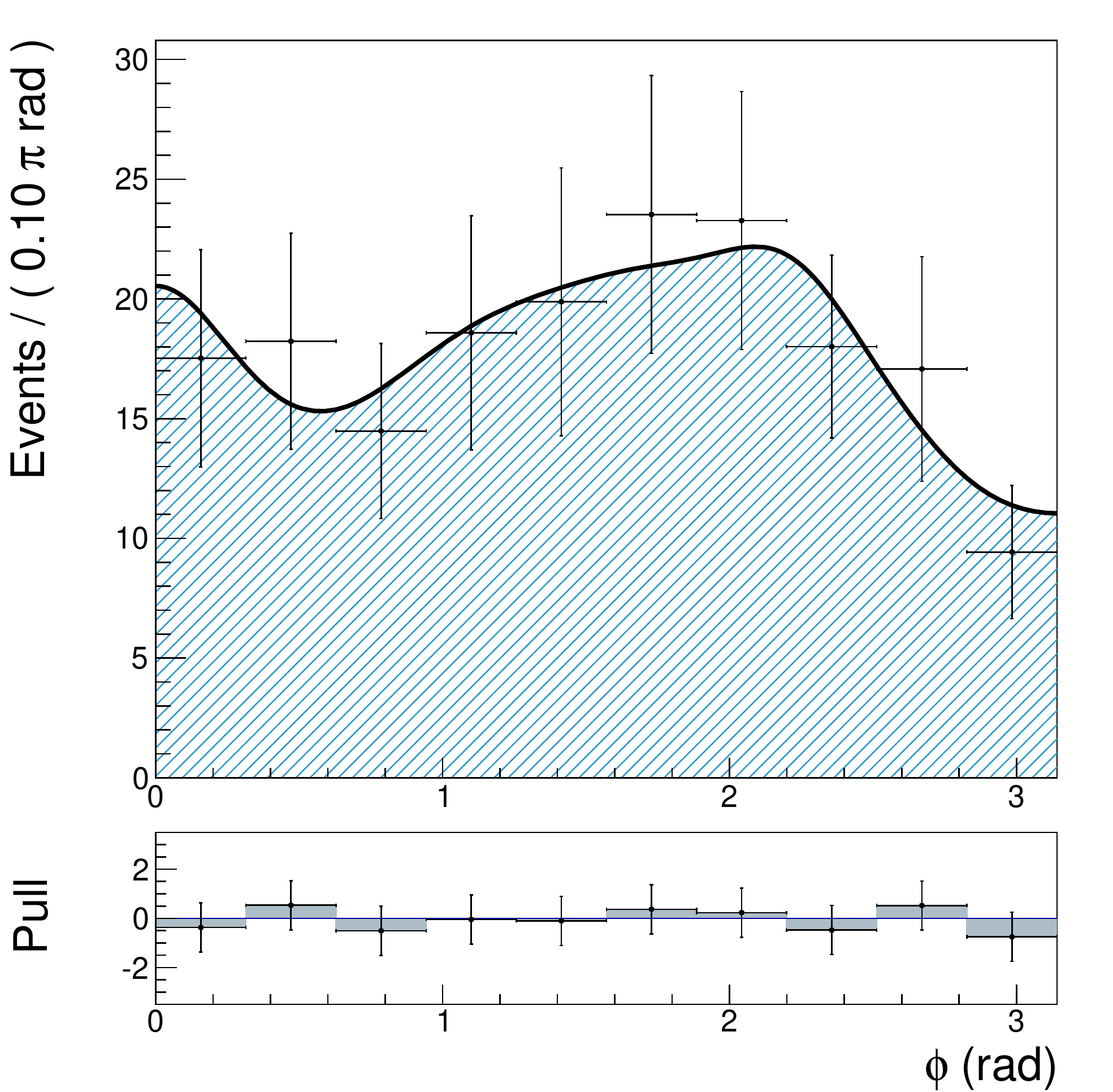}          
    }%
\\
    \subfigure[Projection in $\theta_K$]{
                \includegraphics[width=\factort\textwidth]{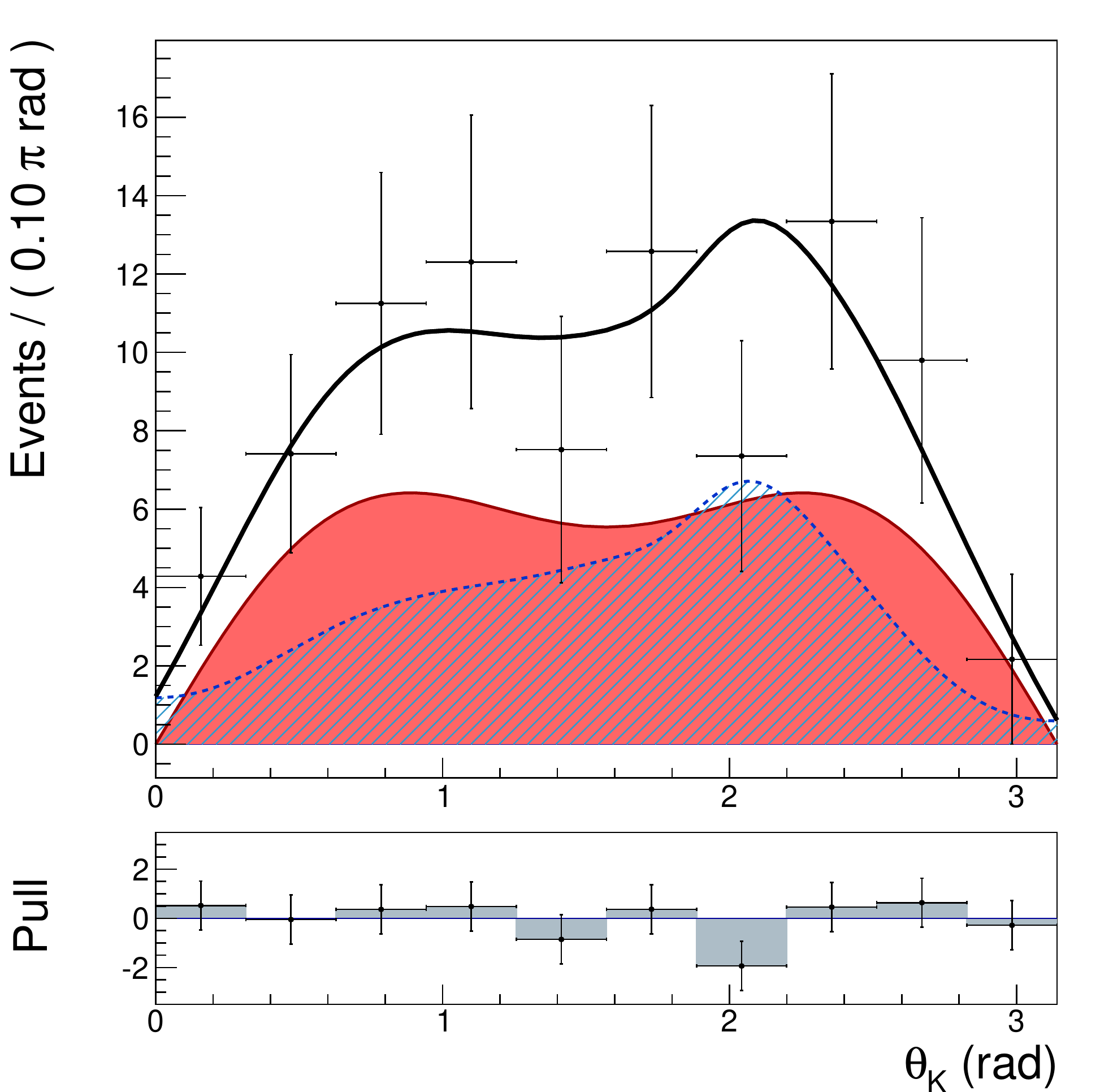}          
    }%
    \subfigure[Projection in $\theta_\ell$]{
                \includegraphics[width=\factort\textwidth]{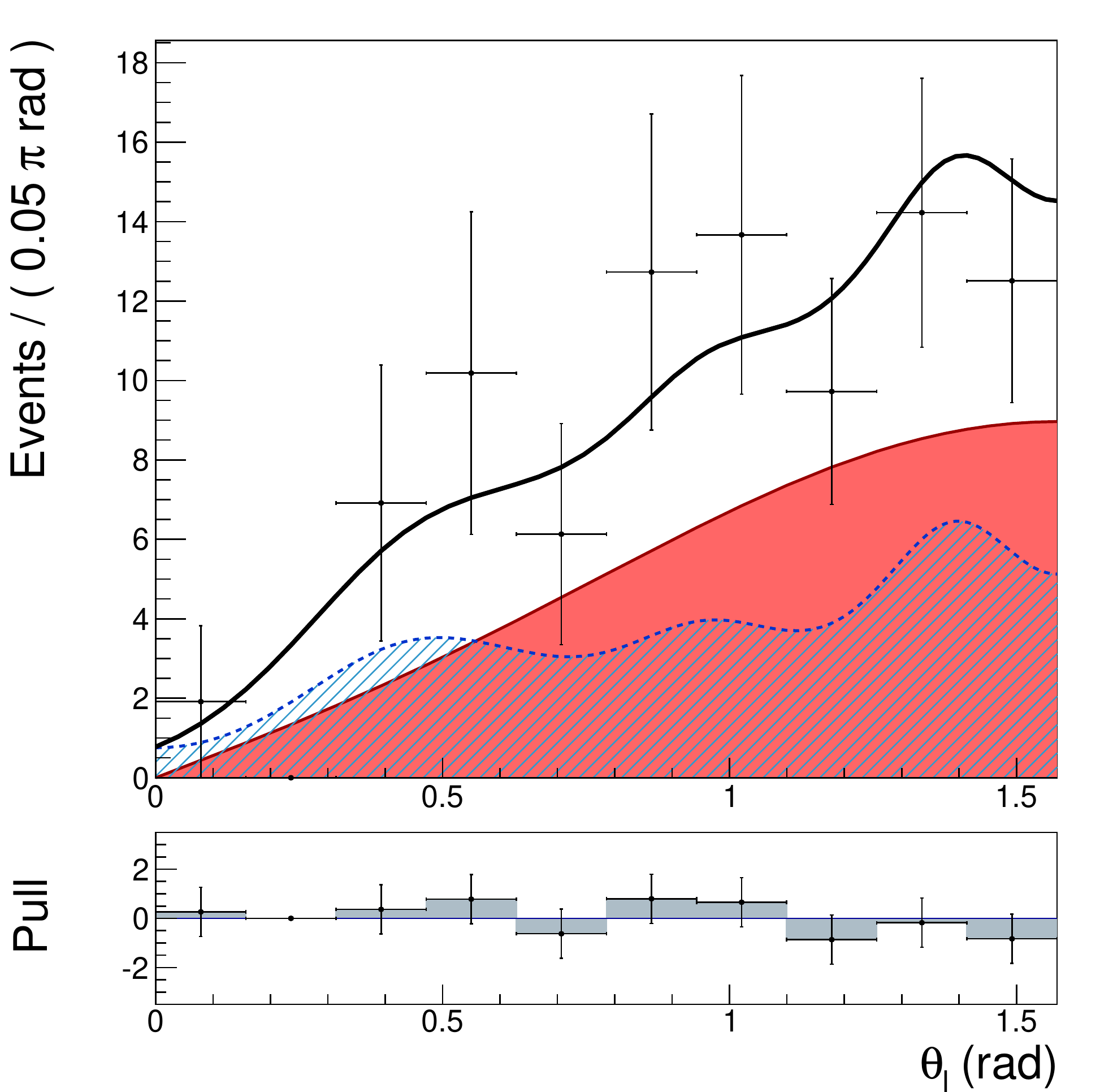}          
    }%
    \subfigure[Projection in $\phi$]{
                \includegraphics[width=\factort\textwidth]{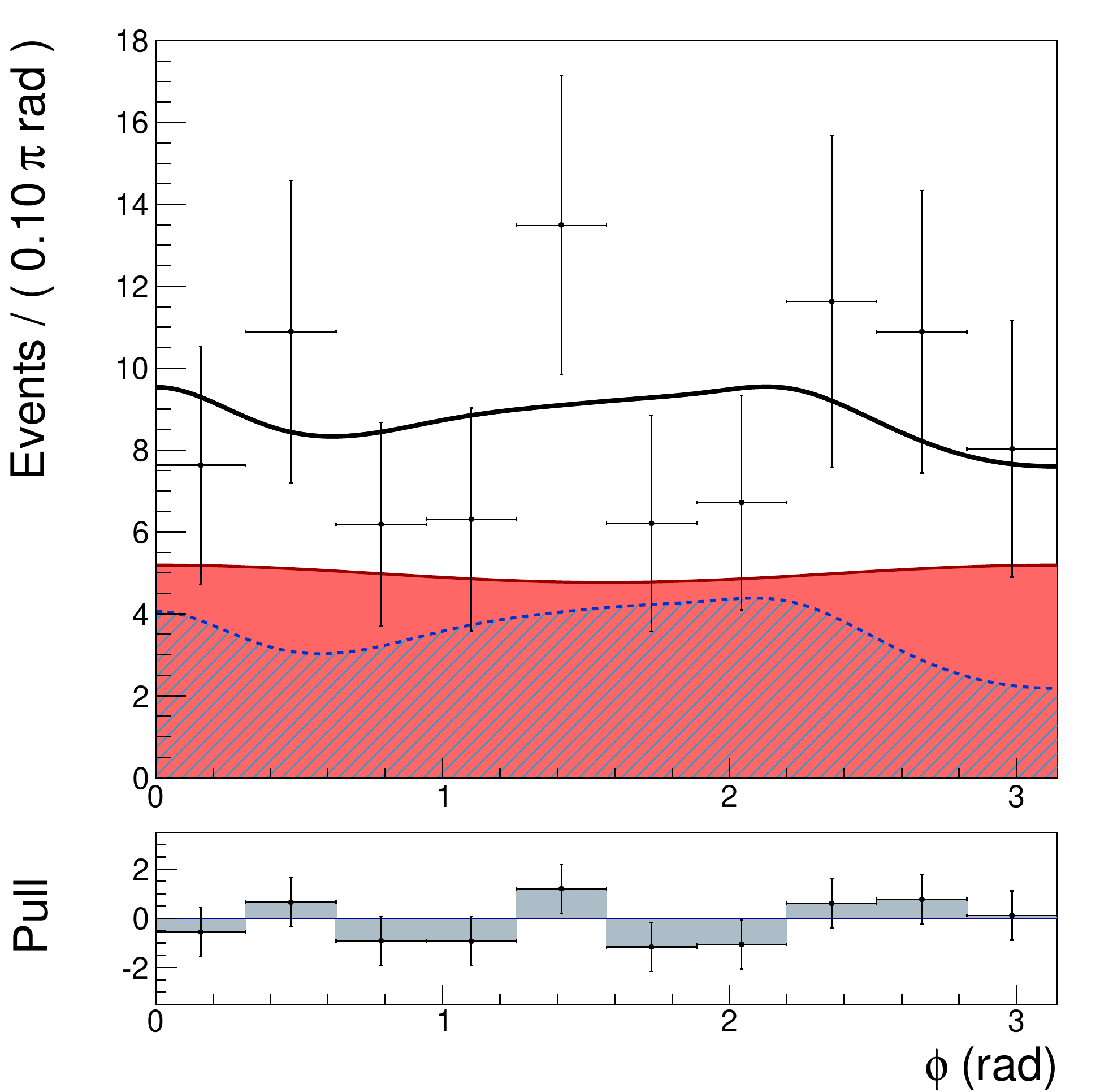}          
    }%
\caption{\fitprojectioncaption{5}{2}}
\label{fig:fit_sample}
\end{figure*}

For $P_5'$ a deviation with respect to the DHMV SM prediction is observed with a significance of $2.1\sigma$  in the $q^2$ range $4.0 < q^2 < 8.0 ~\mathrm{GeV}^2/c^4$.
The  fit result is displayed in Figure \ref{fig:fit_sample} with the corresponding projections.
The distance to the SM prediction from BSZ and JC corresponds to $1.72\sigma$ and $1.68\sigma$, respectively.

The discrepancy in $P_5'$ supports measurements by LHCb \cite{lhcb1}, where a $3.7\sigma$ deviation was observed in the region $4.30 < q^2 < 8.68 ~\mathrm{GeV}^2/c^4$.
To avoid unknown theory errors  originating from the  $J/\psi$ resonance the low $q^2$ region is limited to $q^2<8.0\gevsq$ in the present measurement.
LHCb performed an update on the analysis \cite{lhcb2} with three times the integrated luminosity.
In the update the overall discrepancy of the differential distributions with the DHMV SM prediction was $3.4\sigma$  \cite{lhcb2}.

\begin{table*}
	\centering
	\caption{Results of the angular analysis.  The first errors of the measurement are the statistical and the second the systematic error. Observables are compared to SM predictions  provided by the authors of Refs. \cite{Descotes-Genon:2014uoa,Jager:2012uw,Jager:2014rwa}. }
	\label{tab:angular_results}
	\resizebox{\textwidth}{!}{%
	\begin{tabular}{ @{\hspace{0.5cm}}l c @{\hspace{0.5cm}} r @{\hspace{0.5cm}} r @{\hspace{0.5cm}} r @{\hspace{0.5cm}}r @{\hspace{0.5cm}}}
\hline\hline
$q^2$ in $\mathrm{GeV}^2/c^4$  & Observable & Measurement &  DHMV & BSZ & JC \\
%
%
\hline
$[1.00 , 6.00]$  &  $P_4'$  &  $-0.095^{+0.302}_{-0.309} \pm 0.174$  &   -   & -     &  $-0.300^{+0.090}_{-0.080}$  \\[5pt]
	 &  $P_5'$  &  $\;0.385^{+0.276}_{-0.285} \pm 0.099$  &      -&   -   &  $-0.360^{+0.190}_{-0.170}$  \\[5pt]
	 &  $P_6'$  &  $-0.202^{+0.278}_{-0.270} \pm 0.172$  &    -  &     - &  $0.040^{+0.100}_{-0.100}$  \\[5pt]
	 &  $P_8'$  &  $\;0.440^{+0.311}_{-0.320} \pm 0.195$  &     - &     - &  $0.010^{+0.020}_{-0.019}$  \\[5pt]
\hline
$[0.10 , 4.00]$  &  $P_4'$  &  $\;0.208^{+0.400}_{-0.434} \pm 0.070$  &  $-0.026\pm{0.098}$  &  $-0.029\pm{0.103}$  &  $-0.010^{+0.060}_{-0.060}$  \\[5pt]
	 &  $P_5'$  &  $\;0.631^{+0.403}_{-0.419} \pm 0.067$  &  $0.175\pm{0.086}$  &  $0.199\pm{0.077}$  &  $0.200^{+0.110}_{-0.110}$  \\[5pt]
	 &  $P_6'$  &  $-0.670^{+0.419}_{-0.387} \pm 0.194$  &  $-0.055\pm{0.018}$  &  $-0.056\pm{0.018}$  &  $0.040^{+0.060}_{-0.060}$  \\[5pt]
	 &  $P_8'$  &  $-0.309^{+0.519}_{-0.472} \pm 0.210$  &  $-0.030\pm{0.017}$  &  $-0.031\pm{0.016}$  &  $0.006^{+0.033}_{-0.033}$  \\[5pt]
\hline
$[4.00 , 8.00]$  &  $P_4'$  &  $-0.477^{+0.266}_{-0.252} \pm 0.070$  &  $-0.441\pm{0.106}$  &  $-0.521\pm{0.087}$  &  $-0.490^{+0.070}_{-0.060}$  \\[5pt]
	 &  $P_5'$  &  $-0.267^{+0.275}_{-0.269} \pm 0.049$  &  $-0.881\pm{0.082}$  &  $-0.770\pm{0.100}$  &  $-0.810^{+0.170}_{-0.140}$  \\[5pt]
	 &  $P_6'$  &  $-0.057^{+0.264}_{-0.262} \pm 0.189$  &  $-0.003\pm{0.011}$  &  $-0.002\pm{0.008}$  &  $0.020^{+0.110}_{-0.110}$  \\[5pt]
	 &  $P_8'$  &  $\;0.130^{+0.257}_{-0.259} \pm 0.172$  &  $-0.022\pm{0.010}$  &  $-0.020\pm{0.007}$  &  $0.007^{+0.020}_{-0.021}$  \\[5pt]
\hline
$[10.09 , 12.90]$  &  $P_4'$  &  $-0.088^{+0.414}_{-0.402} \pm 0.114$  &     - &  -    &-      \\[5pt]
	 &  $P_5'$  &  $-0.504^{+0.351}_{-0.327} \pm 0.057$  &    - &     - & -     \\[5pt]
	 &  $P_6'$  &  $-0.341^{+0.347}_{-0.326} \pm 0.222$  &      -&  -    &  -    \\[5pt]
	 &  $P_8'$  &  $-1.017^{+0.415}_{-0.374} \pm 0.207$  &      -&-      &    -  \\[5pt]
\hline
$[14.18 , 19.00]$  &  $P_4'$  &  $-0.371^{+0.272}_{-0.252} \pm 0.074$  &  $-0.632\pm{0.026}$  &  $-0.632\pm{0.026}$  &  -    \\[5pt]
	 &  $P_5'$  &  $-0.547^{+0.257}_{-0.225} \pm 0.058$  &  $-0.601\pm{0.051}$  &  $-0.601\pm{0.051}$  &  -    \\[5pt]
	 &  $P_6'$  &  $\;0.384^{+0.292}_{-0.306} \pm 0.236$  &  $-0.004\pm{0.069}$  &  $-0.004\pm{0.069}$  &  -    \\[5pt]
	 &  $P_8'$  &  $\;0.242^{+0.301}_{-0.309} \pm 0.233$  &  $-0.001\pm{0.028}$  &  $-0.001\pm{0.028}$  &   -   \\[5pt]
%
%
	 \hline\hline
	 \end{tabular}
	 }
\end{table*}

\section{Conclusion}

We present results of the  first angular analysis of \bkllzero in three dimensions at $B$ factories, including both the muon and electron modes.
In total  $117.6\pm12.4$ signal candidates for \Bto{511339} and $69.4\pm 12.0$ signal events for \Bto{511335} are observed.
The signal yields are consistent with those expected from previous measurements.
With the combined data of both channels a full angular analysis in three dimensions in five bins of $q^2$, the di-lepton invariant mass squared, is performed.
A data transformation technique is applied to reduce the dimension of the differential decay rate from eight to three.
By this means the fit is independently sensitive to  observables $P_4'$, $P_5'$, $P_6'$ and $P_8'$, which are optimized regarding theoretical uncertainties from form-factors.
Altogether  20  measurements are performed extracting  $P_{4,5,6}'$ or $P_8'$, the \kast longitudinal polarization     $F_L$ and the transverse polarization asymmetry $A_T^{(2)}$. 
The results are compared with SM  predictions  and  overall agreement is observed.
One measurement is found to deviate by $2.1\sigma$ from the predicted value  into the same  direction  and in the same $q^2$ region where the LHCb collaboration reported the so-called $P_5'$ anomaly \cite{lhcb1,lhcb2}.


\section{Acknowledgments}

We thank J. Virto  and J. Camalich for providing Standard Model predictions for the observables.

We thank the KEKB group for the excellent operation of the
accelerator; the KEK cryogenics group for the efficient
operation of the solenoid; and the KEK computer group,
the National Institute of Informatics, and the 
PNNL/EMSL computing group for valuable computing
and SINET4 network support.  We acknowledge support from
the Ministry of Education, Culture, Sports, Science, and
Technology (MEXT) of Japan, the Japan Society for the 
Promotion of Science (JSPS), and the Tau-Lepton Physics 
Research Center of Nagoya University; 
the Australian Research Council;
Austrian Science Fund under Grant No.~P 22742-N16 and P 26794-N20;
the National Natural Science Foundation of China under Contracts 
No.~10575109, No.~10775142, No.~10875115, No.~11175187, No.~11475187
and No.~11575017;
the Chinese Academy of Science Center for Excellence in Particle Physics; 
the Ministry of Education, Youth and Sports of the Czech
Republic under Contract No.~LG14034;
the Carl Zeiss Foundation, the Deutsche Forschungsgemeinschaft, the
Excellence Cluster Universe, and the VolkswagenStiftung;
the Department of Science and Technology of India; 
the Istituto Nazionale di Fisica Nucleare of Italy; 
the WCU program of the Ministry of Education, National Research Foundation (NRF) 
of Korea Grants No.~2011-0029457,  No.~2012-0008143,  
No.~2012R1A1A2008330, No.~2013R1A1A3007772, No.~2014R1A2A2A01005286, 
No.~2014R1A2A2A01002734, No.~2015R1A2A2A01003280 , No. 2015H1A2A1033649;
the Basic Research Lab program under NRF Grant No.~KRF-2011-0020333,
Center for Korean J-PARC Users, No.~NRF-2013K1A3A7A06056592; 
the Brain Korea 21-Plus program and Radiation Science Research Institute;
the Polish Ministry of Science and Higher Education and 
the National Science Center;
the Ministry of Education and Science of the Russian Federation and
the Russian Foundation for Basic Research;
the Slovenian Research Agency;
Ikerbasque, Basque Foundation for Science and
the Euskal Herriko Unibertsitatea (UPV/EHU) under program UFI 11/55 (Spain);
the Swiss National Science Foundation; 
the Ministry of Education and the Ministry of Science and Technology of Taiwan;
and the U.S.\ Department of Energy and the National Science Foundation.
This work is supported by a Grant-in-Aid from MEXT for 
Science Research in a Priority Area (``New Development of 
Flavor Physics'') and from JSPS for Creative Scientific 
Research (``Evolution of Tau-lepton Physics'').

\bibliography{bib}
\end{document}